\documentclass[letter,twocolumn]{autart}    %

\usepackage{amsmath,amssymb}
\usepackage{graphicx,color,subfigure}
\usepackage{cite}

\usepackage{tikz,calc,bbding}
\usetikzlibrary{shapes, arrows, decorations, decorations.pathreplacing, decorations.pathmorphing, decorations.markings, fit, matrix}
\usepackage{arydshln}
\usepackage{enumitem}
\usepackage[numbered, framed]{mcode}
\usepackage[numbers]{natbib} 

\allowdisplaybreaks

\renewcommand{\theenumi}{(\roman{enumi})}

\graphicspath{{figures/}}

\newtheorem{theorem}{Theorem}[section]
\newtheorem{proposition}[theorem]{Proposition}

\newtheorem{corollary}[theorem]{Corollary}

\newtheorem{definition}[theorem]{Definition}

\newtheorem{problem}{Problem}

\newcommand{\longthmtitle}[1]{\mbox{}{\bf \textit{(#1).}}}

\newcommand{\real}{\ensuremath{\mathbb{R}}}
\newcommand{\realpos}{\ensuremath{\mathbb{R}_{>0}}}
\newcommand{\realnonneg}{\ensuremath{\mathbb{R}_{\ge 0}}}

\DeclareMathOperator*{\argmax}{\arg\max}

\newcommand{\setdef}[2]{\{#1 \; | \; #2\}}
\newcommand{\setdefb}[2]{\big\{#1 \; | \; #2\big\}}

\newcommand{\Uc}{\mathcal{U}}

\newcommand\abf{\mathbf{a}}
\newcommand\bbf{\mathbf{b}}
\newcommand\cbf{\mathbf{c}}

\newcommand\mbf{\mathbf{m}}

\newcommand\ubf{\mathbf{u}}
\newcommand\vbf{\mathbf{v}}

\newcommand\xbf{\mathbf{x}}

\newcommand\Abf{\mathbf{A}}
\newcommand\Bbf{\mathbf{B}}

\newcommand\Ebf{\mathbf{E}}
\newcommand\Fbf{\mathbf{F}}
\newcommand\Gbf{\mathbf{G}}
\newcommand\Ibf{\mathbf{I}}

\newcommand\Nbf{\mathbf{N}}
\newcommand\Pbf{\mathbf{P}}

\newcommand\Rbf{\mathbf{R}}
\newcommand\Tbf{\mathbf{T}}

\newcommand\Wbf{\mathbf{W}}

\newcommand\sigmab{{\boldsymbol{\sigma}}}
\newcommand\Sigmab{\boldsymbol{\Sigma}}

\newcommand\Pib{\boldsymbol{\Pi}}

\newcommand{\ones}{\mathbf{1}}
\newcommand{\zeros}{\mathbf{0}}

\newcommand{\Pbb}{\mathbb{P}}
\newcommand{\Tbb}{\mathbb{T}}

\newcommand{\diag}{{\rm diag}}
\newcommand{\until}[1]{\{1,\dots,#1\}}
\newcommand{\ontil}[1]{\{0,\dots,#1\}}

\newcommand\tr{\text{tr}}

\renewcommand\det{\text{det}}

\newcommand\s{{\rm s}}
\newcommand\zls{\{0, \ell, \s\}}
\newcommand\U{\Uc}
\newcommand{\vect}[1]{\mathbf{#1}}
\newcommand{\vectS}[1]{\boldsymbol{#1}_{\boldsymbol{\sigma}}}

\newcommand{\card}[1]{\lvert#1\rvert}

\newcommand{\norm}[1]{\left\lVert#1\right\rVert}
\newcommand\chireg{\chi_{\rm reg}}
\newcommand\chipp{\chi_{\rm pp}}
\newcommand\chiosc{\chi_{\rm osc}}
\newcommand{\sign}{\text{sign}}

\renewcommand{\footnoterule}{%
  \hspace{3pt} \hrule width 0.4\textwidth height 0.5pt
  \kern 2pt
}

\newcommand{\oprocendsymbol}{\hbox{$\square$}}
\newcommand{\oprocend}{\relax\ifmmode\else\unskip\hfill\fi\oprocendsymbol}

\tikzset{->-/.style={decoration={
  markings,
  mark=at position #1 with {\arrow{>}}},postaction={decorate}}}

\begin{document}

\begin{frontmatter}

  \title{Structural Characterization of Oscillations in Brain Networks
    with Rate Dynamics}

  \thanks{A preliminary version of this paper appeared at the 2019
    American Control Conference as~\citep{EN-JC:19-acc}. During the
    preparation of the bulk of this work, E. Nozari and R. Planas were
    affiliated with the
 University of California, San Diego.}

\author[first]{Erfan Nozari}
\author[second]{\quad Robert Planas}
\author[third]{\quad Jorge Cort\'es}

\address[first]{Department of Mechanical Engineering, University of
  California, Riverside, erfan.nozari@ucr.edu.}

\address[second]{Department of Mechanical and Aerospace Engineering,
  University of California, Irvine, planasr@uci.edu.}

\address[third]{Department of Mechanical and Aerospace Engineering,
  University of California, San Diego, cortes@ucsd.edu.}

\begin{abstract}
  Among the versatile forms of dynamical patterns of activity
  exhibited by the brain, oscillations are one of the most salient and
  extensively studied, yet are still far from being well understood.
  In this paper, we provide various structural characterizations of
  the existence of oscillatory behavior in neural networks using a
  classical neural mass model of mesoscale brain activity called
  linear-threshold dynamics. Exploiting the switched-affine nature of
  this dynamics, we obtain various necessary and/or sufficient
  conditions on the network structure and its external input for the
  existence of oscillations in (i) two-dimensional
  excitatory-inhibitory networks (E-I pairs), (ii) networks with one
  inhibitory but arbitrary number of excitatory nodes, (iii) purely
  inhibitory networks with an arbitrary number of nodes, and (iv)
  networks of E-I pairs. Throughout our treatment, and given the
  arbitrary dimensionality of the considered dynamics, we rely on the
  lack of stable equilibria as a system-based proxy for the existence
  of oscillations, and provide extensive numerical results to support
  its tight relationship with the more standard, signal-based
  definition of oscillations in computational neuroscience.
\end{abstract}

\end{frontmatter}

\section{Introduction}\label{sec:intro}

Oscillations are among some of the first forms of neuronal activity to
be discovered in the human brain, thanks particularly to the invention
of electroencephalogram (EEG) nearly a century
ago~\citep{HB:29}. Thanks to their conceptual simplicity, prominence,
and unmistakable correlation with various neurocognitive processes,
oscillations have since been the subject of significant research from
experimental and computational perspectives in
neuroscience~\citep{GB-AD:04,XW:10,MPJ-TJS:14,PF:15,SRC-BV:17,LP-CWL-DB-DSB:20}. Nevertheless,
the precise mechanisms by which oscillations are generated are still
not understood. In this work, we seek to shed light on this
challenging problem using an analytical, system-theoretic approach and
the linear-threshold mean-field model of neuronal dynamics. Our
results constitute some of the first rigorous characterizations of the
existence of oscillations in these networks, spanning various network
architectures from simple, two-dimensional networks to arbitrarily
complex interconnections of them.

\emph{Literature Review:}
 Oscillations have been the subject of extensive research in the
 neuroscience literature, see,
 e.g.~\citep{GB-AD:04,XW:10,MPJ-TJS:14,PF:15,SRC-BV:17,LP-CWL-DB-DSB:20},
often from solely experimental and/or numerical perspectives. In comparison, \emph{analytical characterizations} of oscillations have remained far behind, even though they can enable a more precise understanding of the
role that different network components and their interconnections
have in the appearance of oscillations, with potential implications for
the study of abnormal behavior (e.g., epilepsy, Parkinson), information
transmission, medical interventions, and beyond.
Among analytical studies, the Wilson-Cowan
model~\citep{HRW-JDC:72} has played a special role owing to its
minimal architecture and richness of non-trivial dynamics at the same
time. Nevertheless, analytical characterization of structural
conditions giving rise to oscillations even in the Wilson-Cowan model has
not moved beyond partial
results~\citep{BB:86,RMB-ABK:92,LHAM-MAB-JGCB:02,ACEO-MWJ-RB:14},
mainly due to the intractability of the sigmoidal nonlinearity in the
standard model.
This has motivated the study of variants of the sigmoidal activation
function, such as linear-threshold models.  Analytical results have
been developed in~\citep{SC-DW:96} for the Wilson-Cowan model with
bounded linear-threshold activation functions, but only under a number
of unrealistic assumptions (notably, the violation of Dale's law,
excluding interaction terms inside the nonlinear activation functions,
and a chain network topology).
Similar to neural mass models with sigmoidal nonlinearities, models
with linear-threshold activation functions%
\footnote{not to be confused with binary networks subject to linear
  integration and thresholding, such as the Hopfield
  network~\citep{JJH:82}.} %
also exhibit rich nonlinear phenomena including multistability, limit
cycles, chaos, and bifurcations, see
e.g.,~\citep{KM-AD-VI-CC:16,FC-AA-FP-JC:21-csl}. In our previous work,
we characterized the existence and uniqueness of equilibria and
asymptotic stability in linear-threshold networks with arbitrary
topologies~\citep{EN-JC:21-tacI} and also provided sufficient and
necessary conditions for the existence of oscillations in
two-dimensional excitatory-inhibitory networks and their networked
excitatory-coupled interconnection~\citep{EN-JC:19-acc}.  Among other
contributions, the present work extends this characterization to the
existence of oscillations in excitatory-inhibitory networks with
arbitrary number of excitatory nodes and networks of two-dimensional
oscillators with more realistic, excitatory-to-all inter-oscillator
connections.

Oscillations have also been studied extensively using bifurcation
theory (particularly the Hopf bifurcation), see e.g.,
\citep{EMI:07,GBE-DHT:10,JAW-TB-ARK:95,MB-JAR-JRT-SR-NM-PAR:06,
ACEO-MWJ-RB:14,JH-BE:15,
TS-YK-MK-KA:17,MS-HB-SO-AT:20} and references therein. However,
a fundamental limitation of these works, and a major difference with
the approach here, is the univariate nature of the former. In other
words, bifurcation analysis is often conducted by fixing all (of the
numerous) network parameters and studying the effect of varying one or
two parameters at a time. In contrast, our global analysis provides
a complete characterization of the set of all parameters that give
rise to oscillations, a set whose boundaries consist of bifurcation
points.
Significant research has been conducted, in the controls and
neuroscience communities alike, to characterize oscillatory dynamics
using models of phase oscillators, the most notable of which being the
Kuramoto model,
see~\citep{MB-SH-AD:10,%
TM-GB-DSB-FP:18} and
references therein. However, while the Kuramoto model has the
advantage of having a smaller (half) state dimension, it is only a
valid approximation to the Wilson-Cowan model in the \emph{weakly
  coupled} regime~\citep{HGS-PW:90,FCH-EMI:12}, where interconnected
oscillators primarily affect each other's phase dynamics and their
amplitude dynamics can be neglected. Moving beyond the weakly
connected regime, amplitude dynamics, particularly saturations and
phase-amplitude coupling~\citep{MJH-EN-BR:19,ACEO-MWJ-RB:14}, become
critical~\citep{GBE-NK:90} and more complex models, such as the full
Wilson-Cowan model are required.

\emph{Statement of Contributions:} Our main contributions are
fourfold, and consist of conditions on the structure of
linear-threshold networks and their inputs that are necessary and/or
sufficient to guarantee the lack of stable equilibria (LoSE).  Since
conditions for the existence of limit cycles in systems with higher
than two dimensions are unknown in general, we use LoSE (which
constitutes the main condition in the Poincar\'e-Bendixson theory for
existence of limit cycles in planar systems) as a proxy for the
existence of oscillations.
First, motivated by the higher abundance and versatility of excitatory
neurons in the mammalian cortex, we provide a necessary and sufficient
condition for LoSE in networks with a single inhibitory but arbitrary
excitatory nodes. We also describe two important consequences of this
result, including a simple, intuitive, and exact characterization of
limit cycles in the Wilson-Cowan model with linear-threshold
nonlinearity, as well as the fact that purely excitatory networks
always have stable equilibria.
Second, purely inhibitory networks have long been known to be able to
generate oscillations, and are often believed to play a central role
in cortical oscillations in the brain. Our second contribution
consists of an extensive study of LoSE in such networks, where we
provide structural necessary conditions on the synaptic connectivity
matrix for LoSE for arbitrary inhibitory networks and a full
characterization of LoSE for pairwise unstable ones. For the latter
case, we provide a graph-theoretic interpretation for the existence of
inputs that induce oscillations in terms of the presence of special
cycles we term \emph{valid}
in the complete graph whose weights are defined in
terms of the synaptic weight matrix.
Next, we study oscillations in networks of multiple brain regions,
each modeled by a simple Wilson-Cowan oscillator. Our third
contribution consists of exact, necessary and sufficient conditions
for LoSE in such networks, when they are coupled either only through
their excitatory nodes or via both excitatory-to-excitatory and
excitatory-to-inhibitory connections.
Finally, we provide extensive numerical evidence that LoSE is indeed a
near necessary and sufficient system-based proxy for the existence of
oscillations, where the latter is often defined based on the power
spectral density of the system's trajectories.
Together, our results provide the first rigorous characterization of
the existence of oscillations in linear-threshold networks with
several different classes of network architectures, along with the
introduction of a novel proxy for oscillatory systems, whose relevance
is of independent interest for the study of arbitrary dynamical
systems.

\section{Problem Formulation}\label{sec:prob-form}

Consider\footnote{Throughout the paper, we employ the following
  notation. $\real$, $\realpos$, and $\realnonneg$ denote the set of
  reals, positive reals, and nonnegative reals, respectively.
  Bold-faced letters are used for vectors and matrices.  $\ones_n$,
  $\zeros_n$, $\zeros_{m \times n}$, and $\Ibf_n$ stand for the
  $n$-vector of all ones, the $n$-vector of all zeros, the $m$-by-$n$
  zero matrix, and the identity $n$-by-$n$ matrix (we omit the
  subscripts when clear from the context).  Given a vector $\xbf$,
  $x_i = (\xbf)_i$ is its $i$th component.  Likewise, $A_{i j}$ refers
  to the $(i, j)$th entry of a matrix $\Abf$.  For block-partitioned
  $\xbf$, $\xbf_i$ refers to the $i$th block of $\xbf$.  For a vector
  $\sigmab' \in \{0, \s\}^n$ and an index $i \in \until{n}$, we say $i
  \in \sigmab'$ if $\sigmab'_i = \s$ and $i \notin \sigmab'$ if
  $\sigmab'_i = 0$. Further, for a (row/column) vector $\xbf$,
  $\xbf_{\sigmab'}$ is its subvector composed of $x_i, i \in \sigmab'$
  and for a matrix $\abf$, $\abf_{i \sigmab'}$ is a row vector
  composed of $a_{ij}, j \in \sigmab'$. Likewise, $\abf_{i, :}$ is the
  $i$'th row of $\abf$ and $\abf_{:, \sigmab'}$ is the submatrix of
  its columns in $\sigmab'$.  For $x \in \real$, $[x]^+ = \max\{x,
  0\}$ and $[x]_0^m = \min\{\max\{x, 0\}, m\}$, which is extended
  entry-wise to $[\xbf]^+$ and $[\xbf]_\zeros^\mbf$.  Given a vector
  $\mbf \in \realpos^n$, $[\zeros, \mbf] = \prod_{i = 1}^n [0, m_i]$.
  For a set $S$, $|S|$ and $S^c$ denotes its cardinality and
  complement. In block representation of vectors and matrices, we use
  compact notations $[\Abf, \Bbf]$, $[\Abf; \Bbf]$, and $\diag(\Abf,
  \Bbf)$ for horizontal, vertical, and diagonal concatenation and
  $\star$ for arbitrary
  blocks. %
  For $a, b \in \real$, $\Uc(a, b)$ denotes the uniform distribution
  over $[a, b]$.  Finally, we let $\Pbb$ denote the set of P-matrices
  (a matrix is a P-matrix if all the principal minors are positive).}
a neuronal network composed of a large number of neurons that
communicate via sequences of spikes. Grouping together neurons with
similar firing rates, under standard assumptions (see, e.g.,~\citep[Ch
7]{PD-LFA:01}), the mean-field dynamics of the network can be
described by the linear-threshold model
\begin{align}\label{eq:blt}
  \tau \dot \xbf(t) = -\xbf(t) + [\Wbf \xbf(t) + \ubf]_\zeros^\mbf,
  \qquad \xbf(0) \in [\zeros, \mbf],
\end{align} 
where $\xbf \in \real^N$ is the state vector with
$x_i$ denoting the average firing rate of the $i$'th neuronal
population, $\Wbf \in \real^{N \times N}$ is the
matrix of average synaptic connectivities, $\ubf \in \real^N$ is the
vector of average external (background) inputs to the populations,
$\mbf \in \realpos^N$ is the vector of average maximum firing rates,
and $\tau > 0$ is the network time constant.
Note that all solutions are bounded as $[\zeros, \mbf]$ is invariant under~\eqref{eq:blt}.

Our previous work~\citep{EN-JC:21-tacI} characterized the existence
and uniqueness of equilibria and asymptotic stability for a variant
of~\eqref{eq:blt} with unbounded activation function ($\mbf = \infty
\cdot \ones_N$), and these results are readily extensible to arbitrary
finite $\mbf$. However, the existence of oscillations in
linear-threshold dynamics is not as well understood. Further, brain
networks often contain interconnections of multiple coupled
oscillators, and our understanding is even smaller about the
oscillatory behavior of interconnections of~\eqref{eq:blt}.  Our goal
is to characterize the relationship between network structure and the
oscillatory behavior observed in linear-threshold dynamics modeling
brain networks.

\begin{problem}\label{prob}
  We seek to answer the following questions for the bounded
  linear-threshold network dynamics~\eqref{eq:blt}:
  \begin{enumerate}
  \item What are neural oscillations? That is, what is an objective
    definition of oscillatory signals and oscillatory systems?
  \item What network structures give rise to oscillations?
  \item What are the structural conditions for the existence of
    oscillations in networked interconnections of multiple oscillatory
    networks?
  \end{enumerate}
\end{problem}

Following common practice in computational
neuroscience~\citep{GB:06,TD-MH-EJP-PV-PS-RG-TN-AHL-JDW-RTK-AS-BV:20},
we here adopt a broad notion of oscillations that includes both
periodic oscillations (limit cycles) and chaotic ones. In the latter
case, a chaotic behavior is oscillatory if its state trajectories are
near-periodic, as captured by next\footnote{Note the similarity
  (relaxing the need for perfect periodicity) as well as the
  difference (requiring near-periodicity here) of this definition with the
  \emph{Yakubovich self-sustained
    oscillations}~\citep{EAT-VAY:89,VIR:07}.}.

\begin{definition}\longthmtitle{Oscillation}\label{def:osc}
  \rm A state trajectory $\xbf(t), t \ge 0$ of~\eqref{eq:blt} is
  oscillatory if
  \begin{enumerate}
  \item its power spectrum contains distinct and pronounced resonance
    peaks; and
  \item it does not asymptotically converge to a constant limit.
    \oprocend
  \end{enumerate}
\end{definition}

Two remarks about Definition~\ref{def:osc} are in order. First,
property (i) is qualitative and fuzzy in nature, as is the notion of
\emph{oscillation}. Different measures can be used to quantify this
property, such as the \emph{regularity index} $\chireg$,
cf.~\ref{app:lose}. Second, the property (ii) is included in the
definition of an oscillation to limit our focus to sustained
(a.k.a. persistent) oscillations and not transient ones. It is
important to note that both types of oscillations are observed in
neuronal dynamics (see, e.g.,
\citep{GB-AD:04,MS:06-neuro,STK-AP-YT-SCM-AAC:18,RQ-JK-ES-NF-RT-ERB-LGC:19}
for sustained and~\citep{SRJ:16,FvE-AJQ-MWW-ACN:18} for transient),
albeit with potentially different underlying dynamical generators. Our
focus here is on the former category in light of the vast literature
on attractor dynamics in biological neuronal
networks~\citep{HKI-LF-SR-KS:19,MM-PP-GM-SC-PD-SF:13,SK-GP-PS-VG-GLB-FL-PLC:19,NT-CC-MVS-MM:19},
while the latter remains an avenue for future research.

The analytical tools in the study of oscillations are generally
limited to 2-dimensional systems (cf. the Poincar\'e-Bendixson
theory~\citep[Ch 3]{LP:00}) or higher-dimensional systems that are
essentially confined to 2-dimensional manifolds (see,
e.g.,~\citep{WG:77,LAS:10}). Thus, throughout the paper, we use
\emph{lack of stable equilibria (LoSE)} \emph{as a proxy for
  oscillations}. In fact, this condition constitutes the main
requirement in the Poincar\'e-Bendixson theory for existence of limit
cycles. In~\ref{app:lose}, we show numerically that this proxy is a
tight characterization of oscillatory dynamics for the
model~\eqref{eq:blt}.

To study the equilibria of~\eqref{eq:blt}, we use its representation
as a switched affine system~\citep{DL:03,MKJJ:03}. It is straightforward to
show~\citep{EN-JC:21-tacI} that $\real^N$ can be decomposed into
$3^N$ switching regions $\{\Omega_\sigmab\}_{\sigmab \in \zls^N}$
defined by
\begin{align*}
  \xbf \in \Omega_\sigmab \Leftrightarrow
  \begin{cases}
    (\Wbf \xbf + \ubf)_i \in (-\infty, 0]; & \forall i \ \ {\rm s.t.}
    \ \ \sigma_i = 0,
    \\
    (\Wbf \xbf + \ubf)_i \in [0, m_i]; & \forall i \ \ {\rm s.t.} \ \
    \sigma_i = \ell,
    \\
    (\Wbf \xbf + \ubf)_i \in [m_i, \infty); & \forall i \ \ {\rm s.t.}
    \ \ \sigma_i = \s,
  \end{cases}
\end{align*}
where $0$, $\ell$, and $\s$ denote a node in inactive, active (linear), and
saturated state, respectively. Thus, \eqref{eq:blt} can be rewritten
in the switched affine form
\begin{align}\label{eq:bltsl}
  \tau \dot \xbf = (-\Ibf + \Sigmab^\ell \Wbf) \xbf + \Sigmab^\ell
  \ubf + \Sigmab^\s \mbf, \qquad 
  \forall \xbf \in \Omega_\sigmab,
\end{align}
where for any $\sigmab \in \zls^N$, $\Sigmab^\ell \in \real^{N \times
  N}$ and $\Sigmab^\s \in \real^{N \times N}$ are diagonal matrices
with entries
\begin{align*}
  \Sigma_{ii}^\ell = \begin{cases}
    1  &\text{if } \sigma_i = \ell, \\
    0 &\text{if } \sigma_i = 0, \s,
\end{cases}
\qquad
\Sigma_{ii}^\s = \begin{cases}
  1  &\text{if }  \sigma_i = \s, \\
  0 &\text{if } \sigma_i = 0, \ell.
\end{cases}
\end{align*} 
Each $\Omega_\sigmab$ then has a
corresponding \emph{equilibrium candidate}
\begin{align}\label{eq:eqc}
\xbf^*_\sigmab = (\Ibf - \Sigmab^\ell \Wbf)^{-1} (\Sigmab^\ell \ubf + \Sigmab^\s \mbf),
\end{align}
and the equilibria of~\eqref{eq:blt} consist of all equilibrium
candidates $\xbf^*_\sigmab$ that belong to their respective switching
regions. Note, in particular, that while the position of the
equilibrium candidates depend on all four of $\Wbf$, $\ubf$, $\mbf$,
and $\sigmab$, their stability is a sole function of $\Wbf$ and
$\sigmab$.

In what follows, we derive exact as well as simplified
characterizations of LoSE for networks with various (and increasingly
more complex) architectures. The network architectures that we study
respect an important property of mammalian cortical networks, known as
Dale's law~\citep{HRW-JDC:72,PD-LFA:01}, according to which each node
has either an excitatory or inhibitory effect on other nodes, but not
both. This means that each column of $\Wbf$ is either nonnegative or
nonpositive, a condition that we follow throughout the paper.

\section{Oscillations in Single Networks}\label{sec:single-network}

We analyze the dynamics~\eqref{eq:blt} and derive conditions on the
network $(\Wbf, \ubf, \mbf)$ giving rise to oscillatory behavior.

\subsection{Excitatory-Inhibitory Networks}\label{sec:ei}

The reciprocal interactions between excitatory and inhibitory
populations of cortical neurons have long been known to be a major
contributor to cortical oscillations~\citep{MPJ-TJS:14}. Arguably, the
simplest scenario with only one excitatory and one inhibitory
populations (each abstracted to one network node) has been the most
popular in theoretical
neuroscience~\citep{JDC-JN-WvD:16}. Interestingly, this coincides with
the fact that LoSE is, under mild conditions, necessary and sufficient
for the existence of almost globally (excluding trajectories starting
at an unstable equilibrium) asymptotically stable limit cycles when $N
= 2$. This two-dimensional case, hereafter called an \emph{E-I pair},
is the celebrated Wilson-Cowan model used in computational
neuroscience for
decades~\citep{HRW-JDC:72,BB:86,RMB-ABK:92,LHAM-MAB-JGCB:02,ACEO-MWJ-RB:14}. Unlike
the standard model with sigmoidal activation functions, however, the
next result shows that a complete characterization of limit cycles can
be obtained for Wilson-Cowan models with bounded linear-threshold
nonlinearities.

\begin{theorem}\longthmtitle{Limit cycles in E-I
    pairs}\label{thm:ei}
  Consider the dynamics~\eqref{eq:blt} with $N = 2$ and
  \begin{align*}
    \Wbf = \begin{bmatrix}
      a & -b
      \\
      c & -d
    \end{bmatrix}
          , \qquad a,
          b, c, d \ge 0.
  \end{align*}
  All network trajectories (except those starting at an unstable
  equilibrium, if any) converge to a limit cycle if and only if
  \begin{subequations}\label{eq:ei}
    \begin{align}
      \label{eq:eia} d + 2 &< a,
      \\
      \label{eq:eib} (a - 1)(d + 1) &< bc,
      \\
      \label{eq:eic} (a - 1) m_1 &< b m_2,
      \\
      \label{eq:eid} 0 < u_1 &< b m_2 - (a - 1) m_1,
      \\
      \label{eq:eie} 0 < (d + 1) u_1 - b u_2 &< \big[bc - (a - 1)(d + 1)\big] m_1.
    \end{align} 
  \end{subequations}
\end{theorem}
\vspace*{-10pt}
\begin{pf}
  By~\citep[Thm 4.1]{SS-KHJ-JL-SS:02}, all the trajectories (except
  those starting at unstable equilibria, if any) converge to a limit
  cycle if and only if the network does not have any stable
  equilibria. This is, nevertheless, not a special case of
  Theorem~\ref{thm:esi} as we here do not presume~\eqref{eq:eia} but
  rather show its necessity together
  with~\eqref{eq:eib}-\eqref{eq:eie}.

  If $a < 1$, then all the regions $\Omega_\sigmab, \sigmab \in
  \zls^2$ are stable, ensuring the existence of a stable equilibrium
  (since the existence of an equilibrium is always guaranteed by the
  Brouwer fixed point theorem~\citep{LEJB:11}). Thus, assume $a \ge
  1$. Then, as shown in the proof of Theorem~\ref{thm:esi}, the
  trivially stable regions $(\sigma', j), \sigma' \in \{0, \s\}, j \in
  \zls$ do not contain their equilibrium candidates iff $u \in
  Y^c$. One can readily show
  \begin{align*}
    Y &= \Big\{(u_1, u_2) \; | \; u_1 \le \max\big\{0, \min\{b m_2,
    \frac{b}{d + 1} u_2\}\big\} \ \text{or}
    \\
    &u_1 \ge -(a - 1) m_1 + \min\big\{b m_2, \max\{0, \frac{b(u_2 + c
      m_1)}{d + 1}\}\big\}\Big\}.
  \end{align*}
  Therefore, $u \in Y^c$ if and only if
  \begin{subequations}\label{eq:yc}
    \begin{align}
      \label{eq:yca} u_1 &> 0,
      \\
      \label{eq:ycb} u_1 &< b m_2 - (a - 1) m_1, 
      \\
      \label{eq:ycc} u_1 &> \min\{b m_2, \frac{b}{d + 1} u_2\}, 
      \\
      \label{eq:ycd} u_1 &< -(a - 1) m_1 + \max\{0, \frac{b(u_2 + c m_1)}{d + 1}\}.
    \end{align}
  \end{subequations}
  For~\eqref{eq:yc} to be feasible, it is necessary and sufficient that
  \begin{subequations}\label{eq:pairs} 
    \begin{align}
      \label{eq:pairsa} &\text{\eqref{eq:yca} and~\eqref{eq:ycb} :} \
      b m_2 - (a - 1) m_1 > 0,
      \\
      \label{eq:pairsb} &\text{\eqref{eq:yca} and~\eqref{eq:ycd} :} \
      u_2 > - \frac{bc - (a - 1)(d + 1)}{b} m_1,
      \\
      \label{eq:pairsc} &\text{\eqref{eq:ycb} and~\eqref{eq:ycc} :} \
      u_2 < \frac{d + 1}{b} (b m_2 - (a - 1) m_1),
      \\
      \label{eq:pairsd} &\text{\eqref{eq:ycc} and~\eqref{eq:ycd} :} \
      bc > (a - 1)(d + 1).
    \end{align}
  \end{subequations} 
  
  Conditions~\eqref{eq:pairsa} and~\eqref{eq:pairsd} are the same
  as~\eqref{eq:eic} and~\eqref{eq:eib}, respectively. Furthermore,
  under~\eqref{eq:pairs}, \eqref{eq:yc} simplifies to~\eqref{eq:eid}
  and~\eqref{eq:eie}, which in turn ensure~\eqref{eq:pairsb}
  and~\eqref{eq:pairsc}. In conclusion, $u \in Y^c$ if and only
  if~\eqref{eq:eib}-\eqref{eq:eie} hold.

  What remains to study are the regions $(\ell, 0)$, $(\ell, \s)$, and
  $(\ell, \ell)$. The first two are not stable since $a \ge 1$. Also,
  though not needed, they do not include their equilibrium candidates
  due to~\eqref{eq:eid}. On the other hand, for $\sigmab = (\ell,
  \ell)$, 
  \begin{align*}
    \xbf^*_\sigmab = \frac{1}{bc - (a - 1)(d + 1)} \begin{bmatrix} (d
      + 1) u_1 - b u_2 \\ c u_1 - (a - 1) u_2 \end{bmatrix} = \Wbf
    \xbf^*_\sigmab + \ubf.
  \end{align*}
  The first component of $\Wbf \xbf^*_\sigmab + \ubf$ clearly belongs
  to $[0, m_1]$ by~\eqref{eq:eib} and~\eqref{eq:eie}. For its second
  component, we have%
  \footnote{We assume $a \neq 1$ since $(\Wbf \xbf^*_\sigmab +
    \ubf)_2 \in [0, m_2]$ trivially if~$a = 1$.}
  \begin{align*}
    \eqref{eq:eib}, \eqref{eq:eie} &\Rightarrow c u_1 > (a - 1) u_2,
    \\
    \eqref{eq:eid}, \eqref{eq:eie} &\Rightarrow u_2 > \frac{c}{a - 1} u_1
    - \frac{bc - (a - 1)(d + 1)}{a - 1} m_2,
  \end{align*}
  ensuring that $\sigmab = (\ell, \ell)$ always contains its
  equilibrium candidate. Therefore, this region must be unstable
  which, under~\eqref{eq:eib}, happens if and only if $a > d + 2$.
  \qed
\end{pf}

While the simplicity of this two-dimensional E-I model has led to its
long-standing popularity in the computational neuroscience literature,
it clearly comes at the price of limited flexibility to model the
complex dynamics of the brain. In the rest of this paper, we extend
the above analysis to more complex scenarios, beginning with the
following analysis of higher-dimensional excitatory-inhibitory
networks.

Inhibitory neurons constitute about $20\%$ of neurons in the cortex
and have broader (less specific) interconnection and activity patterns
than excitatory neurons. Therefore, we focus on networks with a single
inhibitory node and arbitrary number of excitatory nodes.  Let $N = n
+ 1$, $n \ge 1$, and consider
\begin{align}\label{eq:esi}
  \Wbf = 
  \begin{bmatrix}
    \abf & -\bbf
    \\
    \cbf & -d 
  \end{bmatrix},
  \quad \ubf = 
  \begin{bmatrix}
    \ubf_e
    \\
    u_{n+1}
  \end{bmatrix},
  \quad \mbf =
  \begin{bmatrix}
    \mbf_e
    \\
    m_{n+1} 
  \end{bmatrix} ,
\end{align}
where $\abf \in \realnonneg^{n \times n}$, $\bbf \in \realnonneg^{n
  \times 1}$, $\cbf \in \realnonneg^{1 \times n}$, $d \in
\realnonneg$. Note that this class of networks includes, as a special
case, the well-known 2-dimensional Wilson-Cowan model ($n = 1$)
extensively used in the computational
neuroscience~\citep{AD-TJS:09}.%
We are ready to give our first result on LoSE
for~\eqref{eq:blt}-\eqref{eq:esi}.

\begin{theorem}\longthmtitle{Networks with a single inhibitory
    node}\label{thm:esi}
  Consider the dynamics~\eqref{eq:blt}, \eqref{eq:esi} and assume that
  \begin{align}\label{eq:agd2}
    a_{ii} > d + 2 \qquad \forall i \in \until{n}.
  \end{align}
  Then, the network does not have any stable equilibria iff $ \ubf \in
  \real^{n+1} \setminus Y$, where
  \begin{align*}
    Y &= \bigcup\nolimits_{\sigmab' \in \{0, \s\}^n} \bigg[\bigcap\nolimits_{i \in
      \sigmab'} \!\left(Y_{\sigmab', \s, i} \cup (Y_{\sigmab', 0, i}
      \cap Y_{\sigmab', \ell, i})\right) \cap
    \\
    &\hspace{69pt}\bigcap\nolimits_{i \notin \sigmab'} \!\left(Y_{\sigmab', 0, i} \cup
      (Y_{\sigmab', \s, i} \cap Y_{\sigmab', \ell, i})\right) \bigg],
      \\
  Y_{\sigmab'\!, j, i} &= \begin{cases}
  \setdef{\ubf}{u_i \ge y_{\sigmab'\!, j, i}} \ ; & \text{\rm if} \ i \in \sigmab' \\
  \setdef{\ubf}{u_i \le y_{\sigmab'\!, j, i}} \ ; & \text{\rm if} \ i \notin \sigmab'
  \end{cases} \quad \forall j \in \zls,
  \end{align*} 
  $y_{\sigmab'\!, 0, i} = -(\abf_{i \sigmab'} \!-\! \Ibf_{i \sigmab'})
  \mbf_{\sigmab'}$, $y_{\sigmab'\!, \s, i} = y_{\sigmab'\!, 0, i} +
  b_i m_{n + 1}$, and $y_{\sigmab'\!, \ell, i} = y_{\sigmab'\!, 0, i}
  + \frac{b_i (u_{n + 1} + \cbf_{\sigmab'} \mbf_{\sigmab'})}{d + 1}$
  for $\sigmab' \in \{0, s\}^n$ and $i \in \until{n}$.
\end{theorem}
\vspace*{-10pt}
\begin{pf}
  The proof consists of two steps: first, we determine the list of
  $\Omega_\sigmab$ that are stable and second, we ensure that they do
  not contain their equilibrium candidates iff $ \ubf \in \real^{n+1}
  \setminus Y$.

  \emph{Step 1:} The switching regions can be naturally decomposed
  into two groups: those in which at least one of the excitatory nodes
  is active and those in which all the excitatory nodes are either
  inactive or saturated. We next show that these correspond to
  unstable and stable switching regions, respectively. Consider any
  $\sigmab \in \zls^N$ and let $L = \setdef{i \in \until{n}}{\sigma_i
    = \ell}$ (note that $L$ is independent of $\sigma_{n + 1}$). Let
  $r = |L|$, and let $\Pib$ be the permutation matrix such that
  $\Pib \sigmab = (\zeros_{n - r}, \ell, \dots, \ell, \sigma_{n +
    1})$.  The coefficient matrix $-\Ibf + \Sigmab \Wbf$ in the region
  $\Omega_\sigmab$ then satisfies $\Pib (-\Ibf + \Sigmab \Wbf) \Pib^T
  = [-\Ibf_{n - r}, \zeros; \star, \Pbf]$, where $\Pbf = [-\Ibf_r +
  \abf_L, \star; \star, -1 - \Sigma_{n + 1, n + 1} d]$, $\abf_L$ is
  the principal submatrix of $\abf$ composed of its rows and columns
  in $L$, and $\Sigma_{n + 1, n + 1}$ is the bottom-right element of
  $\Sigmab$. Thus, the eigenvalues of $-\Ibf + \Sigmab \Wbf$ consist
  of $(-1)$ with multiplicity $n - r$ and the eigenvalues of
  $\Pbf$. Therefore,
  \begin{itemize}
  \item if $r > 0$, $\Omega_\sigmab$ is unstable since
    $\tr(\Pbf) = \sum_{i \in L} (a_{ii} - 1) - 1 - \Sigma_{n + 1, n
        + 1} d \ge \sum_{i \in L} (a_{ii} - 1) - 1 - d
      \stackrel{\eqref{eq:agd2}}{>} 0$;
  \item if $r = 0$, $\Omega_\sigmab$ is stable since $P = -1 -
    \Sigma_{n + 1, n + 1} d < 0$.
  \end{itemize}

  \emph{Step 2:} According to Step 1, we only need to ensure that
  regions $\Omega_\sigmab$ with $r = 0$ do not contain their
  equilibrium candidates.%
  \footnote{Note that if an equilibrium lies at the boundary of a
    stable switching region, it still attracts (at least half of)
    nearby trajectories: if all the switching regions sharing an
    equilibrium are stable, their coefficient matrices $\{\!-\Ibf +
    \Sigmab \Wbf\} \!\subseteq\! \{-\Ibf, [-\Ibf, \zeros; \cbf,
    -\!1\!-d]\}$ hence share a common quadratic Lyapunov function. If
    an equilibrium is also shared with an unstable switching region,
    it is not difficult to show that the switching hyperplane between
    the stable and unstable regions coincides with the slow eigenspace
    of the coefficient matrices $\{-\Ibf + \Sigmab \Wbf\}$ of the
    stable regions, ensuring that the equilibrium attracts all
    trajectories initiating in the stable side.} %
  These regions have the form
  \begin{align*}
    \sigmab = (\sigmab', j), \qquad \sigmab' \in \{0, \s\}^n, \ {j}
    \in \{0, \ell, \s\}.
  \end{align*}
  We consider three cases based on the value of $j$.
  \begin{enumerate}
  \item $j = 0$: It is straightforward to verify that
    \begin{align*}
      \Wbf \xbf^*_\sigmab + \ubf = 
      \begin{bmatrix}
        \abf_{:, \sigmab'} \mbf_{\sigmab'} + \ubf_e
        \\
        \cbf_{\sigmab'} \mbf_{\sigmab'} + u_{n + 1} 
      \end{bmatrix},
    \end{align*}
    and that $\Wbf \xbf^*_\sigmab + \ubf \in \Omega_\sigmab$ if and
    only if $\ubf \in \bigcap_{i = 1}^{n + 1} Y_{\sigmab', 0, i}$
    where $Y_{\sigmab', 0, n + 1} = \setdef{\ubf}{u_{n + 1} \le
      -\cbf_{\sigmab'} \mbf_{\sigmab'}}$.
  \item $j = \s$: similarly, it follows that
    \begin{align*}
      &\Wbf \xbf^*_\sigmab + \ubf = \begin{bmatrix} \abf_{:, \sigmab'}
        \mbf_{\sigmab'} - \bbf m_{n + 1} + \ubf_e \\ \cbf_{\sigmab'}
        \mbf_{\sigmab'} - d m_{n + 1} + u_{n + 1} \end{bmatrix},
      \end{align*}
      and $\Wbf \xbf^*_\sigmab + \ubf \in \Omega_\sigmab \Leftrightarrow
      \ubf \in \bigcap_{i = 1}^{n + 1} Y_{\sigmab', \s, i}$
      where $Y_{\sigmab', \s, n + 1} = \setdef{\ubf}{u_{n + 1} \ge (d + 1)
        m_{n + 1} - \cbf_{\sigmab'} \mbf_{\sigmab'}}$.
  \item $j = \ell$: it also follows similarly that
    \begin{align*}
      &(d + 1) \Wbf \xbf^*_\sigmab + \ubf =
      \\
      &\begin{bmatrix} (\abf_{:, \sigmab'} (d + 1) - \bbf
        \cbf_{\sigmab'}) \mbf_{\sigmab'} - \bbf u_{n + 1} + (d + 1)
        \ubf_e \\ \cbf_{\sigmab'} \mbf_{\sigmab'} + u_{n +
          1} \end{bmatrix},
      \end{align*}
      and $\Wbf \xbf^*_\sigmab + \ubf \in \Omega_\sigmab
      \Leftrightarrow \ubf \in \bigcap_{i = 1}^{n + 1} Y_{\sigmab',
        \ell, i}$ where $Y_{\sigmab', \s, n + 1} =
      \setdef{\ubf}{-\cbf_{\sigmab'} \mbf_{\sigmab'} \le u_{n + 1} \le
        (d + 1) m_{n + 1} - \cbf_{\sigmab'} \mbf_{\sigmab'}}$.
  \end{enumerate}
  Therefore, for no stable region to contain its equilibrium candidate
  it is necessary and sufficient that
  \begin{align}\label{eq:ybar}
    \ubf \in \real^{n + 1} \setminus \bar Y, \qquad \bar Y =
    \bigcup_{\sigmab' \in \{0, \s\}^n} \bigcup_{j \in \zls} \bigcap_{i
      = 1}^{n + 1} Y_{\sigmab', j, i}.
  \end{align}
  It only remains to show $\bar Y = Y$. For
  $\sigmab' \in \{0, \s\}^n$, let
  \begin{align}\label{eq:ybarsigma}
    \bar Y_{\sigmab'} = \bigcup_{j \in \zls} \bigcap_{i = 1}^{n + 1}
    Y_{\sigmab', j, i}.
  \end{align}
  Then, we have
  $\bar Y_{\sigmab'}^c = \bigcap_{j \in \zls} \bigcup_{i = 1}^{n +
      1} Y_{\sigmab', j, i}^c = \bigcap_{j = 1}^5 (A_j^c \cup B_j^c)$,
  where (in what follows, $^\circ$ denotes the interior of a set)
  \begin{align*}
    A_1 &= \bigcap_{i = 1}^n \!Y_{\sigmab', 0, i}, && B_1 =
    Y_{\sigmab', 0, n + 1}^\circ,
    \\
    A_2 &= A_1, && B_2 = Y_{\sigmab', 0, n + 1} \cap Y_{\sigmab',
      \ell, n + 1},
    \\
    A_3 &= \bigcap_{i = 1}^n \!Y_{\sigmab', \ell, i}, && B_3 =
    Y_{\sigmab', \ell, n + 1}^\circ,
    \\
    A_4 &= A_3, && B_4 = Y_{\sigmab', \ell, n + 1} \cap Y_{\sigmab',
      \s, n + 1},
    \\
    A_5 &= \bigcap_{i = 1}^n \!Y_{\sigmab', \s, i}, && B_5 =
    Y_{\sigmab', \s, n + 1}^\circ,
  \end{align*}
  Since the sets $\{B_j\}_{j = 1}^5$ partition $\real^{n + 1}$, it
  follows 
  that $\bar Y_{\sigmab'}^c = \bigcup_{j = 1}^5 (A_j^c \cap B_j)$, or
  \begin{align}\label{eq:ybarsigmac}
    \notag \bar Y_{\sigmab'}^c &= \bigcup\nolimits_{j \in \{0, \ell, \s\}}
    \Big(\! \Big( \bigcup\nolimits_{i = 1}^n Y_{\sigmab', j, i}^c \Big) \cap
      Y_{\sigmab', j, n + 1} \Big)
    \\
    &= \bigcup\nolimits_{i = 1}^n \bigcup\nolimits_{j \in \{0, \ell, \s\}} (Y_{\sigmab',
      j, i}^c \cap Y_{\sigmab', j, n + 1}).
  \end{align}
  For any $i \in \sigmab'$, we have
  \begin{align}\label{eq:ybarsigmaci1}
    \notag &\bar Y_{\sigmab', i}^c \triangleq \bigcup\nolimits_{j \in \{0,
      \ell, \s\}} (Y_{\sigmab', j, i}^c \cap Y_{\sigmab', j, n + 1})
    \\
    \notag &\stackrel{\text{(a)}}{=} \bigcup\nolimits_{j \in \{0, \ell, \s\}}
    [(Y_{\sigmab', j, i}^c \cap Y_{\sigmab', j, n + 1}) \cup
    (Y_{\sigmab', 0, i}^c \cap Y_{\sigmab', j, n + 1})]
    \\
    \notag &\stackrel{\text{(b)}}{=} \bigcup\nolimits_{j \in \{0, \ell, \s\}}
    (Y_{\sigmab', j, i}^c \cap Y_{\sigmab', j, n + 1}) \cup
    Y_{\sigmab', 0, i}^c
    \\
    \notag &= Y_{\sigmab', 0, i}^c \cup (Y_{\sigmab', \ell, i}^c \cap
    Y_{\sigmab', \ell, n + 1}) \cup (Y_{\sigmab', \s, i}^c \cap
    Y_{\sigmab', \s, n + 1})
    \\
    \notag &\stackrel{\text{(c)}}{=} Y_{\sigmab', 0, i}^c \cup
    (Y_{\sigmab', \ell, i}^c \cap Y_{\sigmab', \s, n + 1}^c) \cup
    (Y_{\sigmab', \s, i}^c \cap Y_{\sigmab', \s, n + 1})
    \\
    \notag &\stackrel{\text{(d)}}{=} Y_{\sigmab', 0, i}^c \cup
    (Y_{\sigmab', \ell, i}^c \cap Y_{\sigmab', \s, n + 1}^c \cap
    Y_{\sigmab', \s, i}^c)
    \\
    \notag &\qquad \qquad \cup (Y_{\sigmab', \s, i}^c \cap
    Y_{\sigmab', \s, n + 1} \cap Y_{\sigmab', \ell, i}^c)
    \\
    \notag &= Y_{\sigmab', 0, i}^c \cup (Y_{\sigmab', \ell, i}^c \cap
    Y_{\sigmab', \s, i}^c)
    \\
    &\stackrel{\text{(d)}}{=} (Y_{\sigmab', 0, i}^c \cup Y_{\sigmab',
      \ell, i}^c) \cap Y_{\sigmab', \s, i}^c,
  \end{align}
  where (a) is because $Y_{\sigmab', 0, i}^c \cap Y_{\sigmab', j, n +
    1} \subseteq Y_{\sigmab', j, i}^c \cap Y_{\sigmab', j, n + 1}$ for
  both $j = \ell$ and $j = \s$ (and is trivial for $j = 0$), (b) is
  because $\{Y_{\sigmab', j, n + 1}\}_{j \in \zls}$ cover $\real^{n +
    1}$, (c) is because $Y_{\sigmab', \ell, i}^c \cap Y_{\sigmab',
    \ell, n + 1} \subseteq Y_{\sigmab', \ell, i}^c \cap Y_{\sigmab',
    \s, n + 1}^c$ and
  \begin{align*}
    (Y_{\sigmab', \ell, i}^c \cap Y_{\sigmab', \s, n + 1}^c) \setminus
    (Y_{\sigmab', \ell, i}^c \cap Y_{\sigmab', \ell, n + 1}) \subseteq
    Y_{\sigmab', 0, i}^c,
  \end{align*}
  (d) is because $Y_{\sigmab', \ell, i}^c \cap Y_{\sigmab', \s, n +
    1}^c \subseteq Y_{\sigmab', \s, i}^c$ and $Y_{\sigmab', \s, i}^c
  \cap Y_{\sigmab', \s, n + 1} \subset Y_{\sigmab', \ell, i}^c$, and
  (e) is because $Y_{\sigmab', 0, i}^c \subseteq Y_{\sigmab', \s,
    i}^c$. By a parallel argument, it can be shown that for any $i
  \notin \sigmab'$,
  \begin{align}\label{eq:ybarsigmaci2}
    \bar Y_{\sigmab', i}^c = (Y_{\sigmab', \s, i}^c \cup Y_{\sigmab',
      \ell, i}^c) \cap Y_{\sigmab', 0, i}^c.
  \end{align}
  Therefore, \eqref{eq:ybar}-\eqref{eq:ybarsigmaci2} gives $\bar Y =
  Y$, completing the proof. \qed
\end{pf}

\begin{figure}%
\centering
  \includegraphics[width=0.85\linewidth]{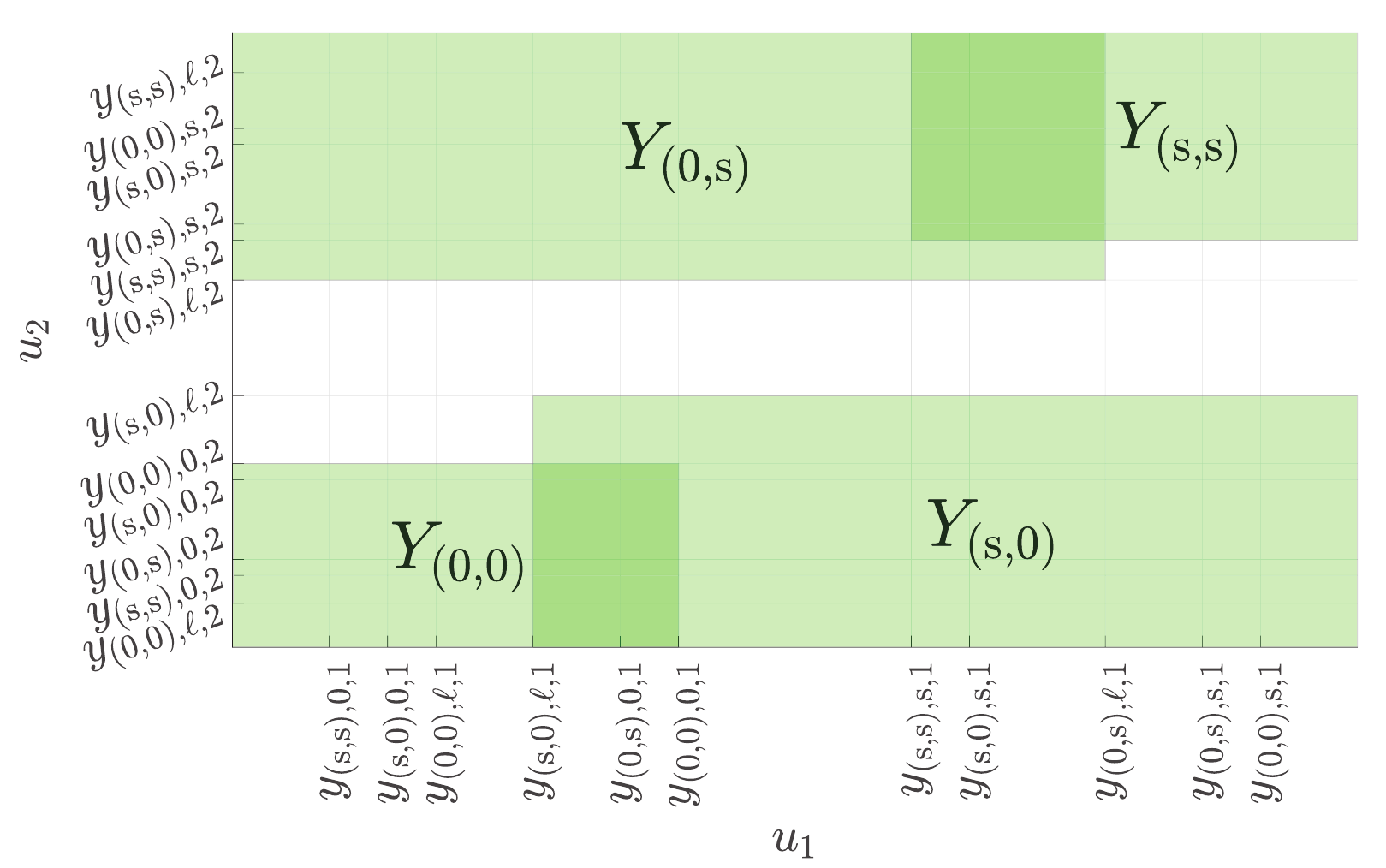} \hspace{20pt}
  \caption{Illustration of the region $Y$ in Theorem~\ref{thm:esi} for
    an example network with $n = 2$. The four shaded quadrants
    represent the cross section of $Y_{\sigmab'}, \sigmab' \in \{0,
    \s\}^2$ at $u_3 = -5$, so the white area is where the network
    lacks any stable equilibria. Network parameters are $\abf
    = [8.5, 1; 1, 5]$, $\bbf = [5; 7]$, $\cbf = [4, 5]$, $d = 1$,
    $\mbf = [2; 3; 6]$.}\label{fig:Y}
  \vspace*{-1ex}
\end{figure}

While the description of $Y$ in Theorem~\ref{thm:esi} may seem
complex, it has a simple interpretation. Consider a fixed value for
$u_{n + 1}$. Then, each of the sets $\left(Y_{\sigmab', \s, i} \cup
  (Y_{\sigmab', 0, i} \cap Y_{\sigmab', \ell, i})\right)$ or
$\left(Y_{\sigmab', 0, i} \cup (Y_{\sigmab', \s, i} \cap Y_{\sigmab',
    \ell, i})\right)$ in the definition of $Y$ are a half space of the
form $\{u_i \ge y\}$ or $\{u_i \le y\}$ (depending on whether $i \in
\sigmab'$ or not) that drive $x_i$ to saturation or inactivity,
respectively.  Therefore, the cross section of $Y$ for this fixed
value of $u_{n + 1}$ is composed of $2^n$ closed orthants, each
unbounded towards a different direction in $\real^n$.
Figure~\ref{fig:Y} shows an example of this for $n = 2$. The union of
these orthants (the shaded area in Figure~\ref{fig:Y}) characterizes
the region where the network has at least one stable equilibrium.

Nevertheless, the set $Y^c$ may in general be non-convex, unbounded,
and disconnected. The next result gives simpler and
easier-to-interpret, but more conservative, conditions. 

\begin{corollary}\longthmtitle{Simpler conditions for networks with a
    single inhibitory node}\label{cor:esi}
  Consider the same assumptions as in Theorem~\ref{thm:esi}. Then, for
  the network not to have any stable equilibria, it is necessary that
  \begin{align}\label{eq:esi-nec}
    -\cbf \mbf_e < u_{n + 1} < (d + 1) m_{n + 1},
  \end{align}
  and sufficient that either
  \begin{subequations}\label{eq:esi-suf1}
    \begin{align}
      \label{eq:esi-suf1a} &0 \le u_{n + 1} \le (d + 1) m_{n + 1} -
      \cbf \mbf_e,
      \\
      \label{eq:esi-suf1b} &\exists i_0 \ {\rm s.t.} \quad (a_{i_0
        i_0} - 1) (d + 1) < b_{i_0} c_{i_0},
      \\
      \label{eq:esi-suf1c} &\frac{b_{i_0} u_{n + 1}}{d \!+\! 1}
      \!\!<\! u_{i_0} \!\!<\!\! \frac{b_{i_0} (u_{n + 1} \!+\! c_{i_0}
        m_{i_0})}{d \!+\! 1} \!-\! (a_{i_0 i_0} \!\!-\! 1) m_{i_0},
      \\
      \label{eq:esi-suf1d} &u_i \!<\! \frac{b_i u_{n \!+\! 1} \!-\!
        [(\abf_{i, :} \!-\! \Ibf_{i, :})(d \!+\! 1) \!-\! b_i \cbf]^+
        \mbf_e}{d \!+\! 1}, \forall i \neq i_0,
    \end{align}
  \end{subequations}
  or
  \begin{subequations}\label{eq:esi-suf2}
    \begin{align}
      \label{eq:esi-suf2a} &(d \!+\! 1) m_{n + 1} \!-\! \min_i (c_i
      m_i) \!\le\! u_{n \!+\! 1} \le (d \!+\! 1) m_{n + 1} ,
      \\
      \label{eq:esi-suf2b} &\exists i_0 \ {\rm s.t.} \quad (a_{i_0
        i_0} - 1) m_{i_0} < b_{i_0} m_{n + 1},
      \\
      \label{eq:esi-suf2c} &0 < u_{i_0} < b_{i_0} m_{n + 1} - (a_{i_0
        i_0} - 1) m_{i_0},
      \\
      \label{eq:esi-suf2d} &u_i < b_i m_{n + 1} - (\abf_{i, :} -
      \Ibf_{i, :}) \mbf_e, \quad \forall i \neq i_0.
    \end{align}
  \end{subequations}
\end{corollary}
\vspace*{-10pt}
\begin{pf} 
  \emph{First, we prove the sufficiency of the conditions
    in~\eqref{eq:esi-suf1}}, by showing that any $\ubf$ satisfying all
  the conditions in~\eqref{eq:esi-suf1} will not belong to $Y$ as
  defined in Theorem~\ref{thm:esi}. Note that the expression for $Y$ can be
  greatly simplified \emph{if} we can restrict $\ubf$ such that for
  all $\sigmab' \in \{0, \s\}^n$,
  \begin{align}\label{eq:allinl}
    \notag Y_{\sigmab', \s, i} \cup (Y_{\sigmab', 0, i} \cap
    Y_{\sigmab', \ell, i}) &= Y_{\sigmab', \ell, i} \qquad \forall i
    \in \sigmab',
    \\
    Y_{\sigmab', 0, i} \cup (Y_{\sigmab', \s, i} \cap Y_{\sigmab',
      \ell, i}) &= Y_{\sigmab', \ell, i} \qquad \forall i \notin
    \sigmab'
  \end{align}
  Given the definition of the sets $Y_{\sigmab', j, i}$, we can see
  that this holds if for all $\sigmab' \in \{0, \s\}^n$,
  \begin{align}\label{eq:unp1}
    -\cbf_{\sigmab'} \mbf_{\sigmab'} \le u_{n+1} \le (d+1)m_{n+1} -
    \cbf_{\sigmab'} \mbf_{\sigmab'},
  \end{align}
  which gives~\eqref{eq:esi-suf1a} since $\max_{\sigmab'}
  -\cbf_{\sigmab'} \mbf_{\sigmab'} = 0$ and $\min_{\sigmab'}
  -\cbf_{\sigmab'} \mbf_{\sigmab'} = - \cbf \mbf_e$.

  Given~\eqref{eq:allinl}, $\ubf$ will not be in $Y$ if and only if
  for any $\sigmab' \in \{0, \s\}^n$, there exists an $i$ such that
  \begin{subequations}
    \begin{align}
      \label{eq:uibdd-a} u_i < \frac{b_i}{d+1} u_{n+1} + \frac{b_i
        \cbf_{\sigmab'} - (\abf_{i\sigmab'} -
        \Ibf_{i\sigmab'})(d+1)}{d+1} \mbf_{\sigmab'},
    \end{align}
    if $i \in \sigmab'$, or
    \begin{align}
      u_i > \frac{b_i}{d+1} u_{n+1} + \frac{b_i \cbf_{\sigmab'} -
        (\abf_{i\sigmab'} - \Ibf_{i\sigmab'})(d+1)}{d+1}
      \mbf_{\sigmab'},
    \end{align}
  \end{subequations}
  if $i \notin \sigmab'$. Note that these are $2^n$ sets of
  inequalities, where at least one inequality needs to be satisfied
  from each set using only the $n$ variables $u_1, \dots, u_n$. This
  provides us with a choice of which inequality from each set we
  choose to enforce, with any choice imposing $2^n$ upper/lower bounds
  on $u_1, \dots, u_n$. Here, care should be taken to ensure the
  resulting system of inequalities is feasible. For any variable
  $u_i$, as long as the inequalities imposed on it are all either
  lower bounds or upper bounds, a feasible $u_i$ exists. However, any
  lower and upper bounds imposed on the same $u_i$ must be ensured to
  be collectively feasible, in turn putting additional restrictions on
  $\Wbf$ and $\mbf$. With this background in mind, we obtain an
  explicit yet minimally restrictive set of sufficient conditions as
  follows.

  Assume that~\eqref{eq:esi-suf1b} holds. Then, we impose the $i$'th
  inequality corresponding to $\sigmab' = \zeros$ and $\sigmab'_i =
  (0, 0, \dots, \s, \dots, 0)$ where the $\s$ is in the $i$'th
  position. The result will be~\eqref{eq:esi-suf1c}, which is feasible
  by~\eqref{eq:esi-suf1b}. For any other $\sigmab' \neq \zeros,
  \sigmab'_i$, we only impose some (or all) of the upper bound
  inequalities in~\eqref{eq:uibdd-a} for $j \neq i$, which will always
  be feasible without any further restrictions on $\Wbf$ and
  $\mbf$. This leads to potentially multiple upper bounds for each $j
  \neq i$, but all of them are greater than the bound
  in~\eqref{eq:esi-suf1b} and are thus satisfied
  if~\eqref{eq:esi-suf1b} is. This completes the proof of the
  sufficiency of~\eqref{eq:esi-suf1}.

  \emph{Second, we prove the sufficiency of the conditions
    in~\eqref{eq:esi-suf2}} following a similar construction. Here,
  instead of~\eqref{eq:allinl}, we restrict $\ubf$ such that for all
  $\sigmab' \neq \zeros$,
  \begin{align}\label{eq:allins}
    Y_{\sigmab', \s, i} \cup (Y_{\sigmab', 0, i} \cap Y_{\sigmab', \ell,
      i}) &= Y_{\sigmab', \s, i} \quad \forall i \in \until{n},
  \end{align}
  and for $\sigmab' = \zeros$,
  \begin{align}\label{eq:allin0}
    Y_{\sigmab', \s, i} \cup (Y_{\sigmab', 0, i} \cap Y_{\sigmab',
      \ell, i}) &= Y_{\sigmab', 0, i} \quad \forall i \in \until{n}.
  \end{align}
  Similar to~\eqref{eq:unp1}, these will hold if $ -\cbf_{\sigmab'}
  \mbf_{\sigmab'} + (d+1) m_{n+1} \le u_{n+1} \le (d+1) m_{n+1}$, for
  all $\sigmab' \neq \zeros$, which is equivalent
  to~\eqref{eq:esi-suf2a}. Then, similar to the proof
  of~\eqref{eq:esi-suf1}, we assume that there exists at least one $i$
  for which~\eqref{eq:esi-suf2b} holds, and enforce the $i$'th
  inequality for $\sigmab' = \zeros$ and $\sigmab' =
  \sigmab'_i$. These together impose~\eqref{eq:esi-suf2c} on $u_i$,
  whose feasibility requires~\eqref{eq:esi-suf2b}. For any other
  $\sigmab'$, we enforce the $j$'th inequality(s) for some (or all) $j
  \in \sigmab', j \neq i$, which requires $u_j < -(\abf_{j\sigmab'} -
  \Ibf_{j\sigmab'}) \mbf_{\sigmab'} + b_j m_{n+1}$ and satisfied if
  the stronger condition~\eqref{eq:esi-suf2d} holds.

  \emph{Finally, we prove the necessity of~\eqref{eq:esi-nec}} by
  contradiction. Assume, first, that $u_{n+1} \ge (d+1) m_{n+1}$. This
  implies that~\eqref{eq:allins} holds for all $\sigmab' \in \{0,
  \s\}^n$ and all $i$. We can then make a sequential argument as
  follows. Starting from $\sigmab' = \zeros$, we would need at least
  one $i$ such that $u_i > b_i m_{n+1}$. This $i$ can then never
  satisfy $u_i < b_i m_{n+1} - (\abf_{i\sigmab'} - \Ibf_{i\sigmab'})
  \mbf_{\sigmab'}$ for any $\sigmab'$, which means that $\ubf$ cannot
  belong to any $Y_{\sigmab', \s, i}^c$ where $i \in \sigmab'$. To
  simplify the discussion and without loss of generality, assume we
  have chosen $i = 1$. Then, $\ubf$ cannot belong to any $Y_{\sigmab',
    \s, 1}^c$ for any $\sigmab' = (\s, \star, \dots,
  \star)$. Therefore, for $\sigmab' = (\s, 0, \dots, 0)$, we need at
  least one $i \ge 2$ such that $u_i > b_i m_{n+1} - a_{i1}
  m_1$. Again, for simplicity and without loss of generality, assume
  $i = 2$. Then, $\ubf$ cannot belong to any $Y_{\sigmab', \s, 2}^c$
  for any $\sigmab' = (\s, \s, \star, \dots, \star)$. Continuing this
  argument, we will ultimately have to impose lower bounds on all the
  elements of $\ubf_e$, which prevent $\ubf$ from belonging to
  $Y_{\sigmab', \s, i}^c$ for $\sigmab' = (\s, \s, \dots, \s)$ and any
  $i$, ensuring the existence of a stable equilibrium by
  Theorem~\ref{thm:esi}, which is a contradiction. An analogous
  argument shows that $u_{n+1} \le -\cbf \mbf_e$ also leads to a
  contradiction. \qed
\end{pf}

Note the parallelism between~\eqref{eq:ei}
and~\eqref{eq:esi-suf1}-\eqref{eq:esi-suf2}. In fact,
conditions~\eqref{eq:esi-suf1b},
\eqref{eq:esi-suf1c}-\eqref{eq:esi-suf1d}, \eqref{eq:esi-suf2b}, and
\eqref{eq:esi-suf2c}-\eqref{eq:esi-suf2d} are generalizations
of~\eqref{eq:eib}, \eqref{eq:eie}, \eqref{eq:eic}, and \eqref{eq:eid},
respectively. Further, Corollary~\ref{cor:esi} has itself a further
consequence with great neuroscientific value, as given next.

\begin{corollary}\longthmtitle{Fully excitatory networks}\label{cor:alle}
  Given the dynamics~\eqref{eq:blt}, if $\Wbf$ is fully excitatory
  (all entries are non-negative), then the network has at least one
  stable equilibrium.
\end{corollary}
\vspace*{-10pt}
\begin{pf}
  A fully excitatory network corresponds to~\eqref{eq:blt},
  \eqref{eq:esi} with a sufficiently negative $u_{n+1}$ that 
  drives the inhibitory node into negative saturation, effectively
  removing it from the network. This happens if, for all~$t$, $ \cbf
  \xbf_e(t) - d x_{n+1}(t) + u_{n+1} < 0 \Leftarrow u_{n+1} < - \cbf
  \xbf_e(t) \Leftarrow u_{n+1} < - \cbf \mbf_e$, which, by
  Corollary~\ref{cor:esi}, implies at least one stable equilibrium
  exists. \qed
\end{pf}

This result can can also be established using the theory of monotone
systems~\citep{MWH-HS:06,DA-EDS:03}.  Corollary~\ref{cor:alle}
provides a simple and rigorous explanation for the well-known
necessity of inhibitory nodes in brain
oscillations~\citep{MAW-RDT-NK-BE-EHB:00}. On the other hand, the
computational neuroscience literature has long shown the possibility
of oscillatory activity in \emph{purely inhibitory}
networks~\citep{MPJ-TJS:14}, an important class of networks that we
treat next.

\subsection{Inhibitory Networks}

Our focus here is on linear-threshold network models~\eqref{eq:blt}
where only inhibitory nodes are present. Consequently,
\begin{equation*}
  \Wbf = 
  \begin{bmatrix}
    - d_{1,1} & -d_{1,2} & \dots & -d_{1,N}
    \\
    -d_{2,1} & -d_{2,2} & \dots & -d_{2,N}
    \\
    \vdots & \ddots & \dots & \vdots
    \\
    -d_{N,1} & -d_{N,2} & \dots & -d_{N,N}
  \end{bmatrix}
\end{equation*}
with $d_{i,j} \ge 0$ for all $i,j$.

\subsubsection{Necessary Conditions for LoSE}
We start by identifying a necessary condition for the lack of stable
equilibria of fully inhibitory networks.

\begin{theorem}\longthmtitle{Necessary condition 
    for oscillatory behavior in fully inhibitory
    networks}\label{thm:NC_FIN_OB}
  If a fully inhibitory network does not have any stable equilibria,
  then $ \Ibf - \Wbf \notin \Pbb$.
\end{theorem}
\vspace*{-10pt}
\begin{pf}
  We argue the counter positive: if $ \Ibf - \Wbf \in \Pbb$, then a
  stable equilibrium point exists. The fact that an equilibrium exists
  for any $\ubf \in \real^N$ is a direct consequence of \citep[Theorem
  IV.1]{EN-JC:21-tacI}. To show it is stable, let us consider any
  switching region $\vectS{\Omega}$ containing the equilibrium.
  Over this region, the dynamics is described by $ -\vect{I} +
  \vect{\Sigma} \Wbf$. Let $r$ be the cardinality of the set of nodes
  in linear state and let $\vect{\Pi}$ be a permutation matrix such
  that $\vect{\Pi} \vect{\sigma} = (\sigma_1,\dots,\sigma_{n-r},
  l,\dots,l)$, where $\sigma_i \in \{0,s\}$.  Then,
  \begin{align}\label{eq:W-decomp-P}
    \vect{\Pi} (-\vect{I} + \vect{\Sigma} \Wbf) \vect{\Pi} ^T
    = \begin{bmatrix} 
      - \vect{I} & 0 \\
      \ast & \vect{P}
    \end{bmatrix},
  \end{align}
  for some matrix $\vect{P}$.  The eigenvalues of the system are
  therefore $-1$, with multiplicity $n-r$, and the eigenvalues
  of~$\Pbf$. Note that $\Pbf$ is a principal submatrix of the
  matrix~$-\vect{I} + \Wbf$.  Since any principal submatrix of a
  ${P}-$matrix is also a $P-$matrix, we deduce $-\Pbf$ is a
  ${P}-$matrix too. In addition, since $\Wbf$ corresponds to a
  fully inhibitory network, $-\Pbf$ is a sign-symmetric matrix,
  meaning that $(-\Pbf(I,J))(-\Pbf(I,J)) \geq 0$ for all $I$,$J
  \subset \lbrace 1, \dots, n \rbrace$ such that $\card{I} =
  \card{J}$.  By \citep[Theorem 1]{AKT-AS-AO-DA:06}, a sign-symmetric
  $P-$matrix is positive stable (if a matrix $\Abf$ is positive
  stable, then $-\Abf$ is stable in the traditional Lyapunov sense, so
  all the eigenvalues of $-\Abf$ have negative real parts).
  Consequently, the eigenvalues of $\Pbf$ fall in the negative complex
  quadrant and the equilibrium is stable. \qed
\end{pf}

Theorem~\ref{thm:NC_FIN_OB} provides a necessary condition based on
the intrinsic properties of the network connectivity. The next result
provides an alternative, much simpler necessary condition based on the
number of nodes.  This result can also be derived using the theory of
monotone systems~\citep{MWH-HS:06,DA-EDS:03}, but we here present an
independent proof that is instructive in the context of our
methodology.

\begin{proposition}\longthmtitle{2-node fully inhibitory networks
    always have a stable equilibrium}\label{prop:2_N_FIN}
  A fully inhibitory network with only two nodes always has a stable
  equilibrium.  
\end{proposition}
\vspace*{-10pt}
\begin{pf}
  We divide the proof in two cases depending on whether
  $(d_{1,1}+1)(d_{2,2}+1) - d_{1,2}d_{2,1} $ is (i) greater than $0$
  or (ii) less than or equal to $0$.  In case (i), the fact that the
  network is fully inhibitory results in all the principal minors of
  $\vect{I} - \Wbf$ being greater than zero, and hence $\vect{I} -
  \Wbf \in \Pbb$.  By Theorem~\ref{thm:NC_FIN_OB}, a stable
  equilibrium point exists.
    
  In case (ii), we look at the equilibrium candidates. Note that only
  one switching region has a non-stable equilibrium candidate (the one
  where both nodes are found in linear state), while all the other
  switching regions have stable equilibrium candidates. Hence, proving
  the existence of multiple equilibrium points in the system is enough
  to prove the stability of it.  By Brouwer's Fixed-Point
  Theorem~\citep{LEJB:11}, an equilibrium point exists. Then, since
  (ii) implies that $\vect{I} - \Wbf \notin \Pbb$, we use
  \citep[Theorem VI.1]{EN-JC:21-tacI} to conclude that the equilibrium
  is not unique. As, at least, two equilibrium points exist, one
  necessarily corresponds to a stable equilibrium candidate. \qed
\end{pf}

\subsubsection{Sufficient Conditions for LoSE}
In the following, we derive sufficient conditions for LoSE by
investigating the instability properties of the equilibrium candidate
of each switching region.
In our study, we focus on the following class of
network structures.

\begin{definition}\longthmtitle{Pairwise unstable network}
  A network $\Wbf$ is pairwise unstable if
  the system matrix $-\vect{I} + \vect{\Sigma} \Wbf$ corresponding
  to each switching region $\vectS{\Omega}$ involving only two nodes
  in linear state is unstable.
\end{definition}

The definition is valid for arbitrary (i.e., not necessarily
inhibitory) networks.  For inhibitory networks, it is equivalent to
asking each principal minor $M_{i,j}$ of order two of $-\vect{I} +
\Wbf$ to be negative, $M_{i,j} < 0$.  Interestingly, this property
allows us to establish conclusions about the instability of the
switching regions that involve more than two nodes in linear state.

\begin{theorem}\longthmtitle{Instability of pairwise unstable 
    networks}\label{PW_I_FUN}
  Let $\Wbf$ be a pairwise unstable
  network. Then, the system matrix $-\vect{I} + \vect{\Sigma}
  \Wbf$ corresponding to each switching region $\vectS{\Omega}$
  involving more than two nodes in linear state is unstable.
\end{theorem}
\vspace*{-10pt}
\begin{pf}
  Let $\vectS{\Omega}$ be a switching region involving more than two
  nodes in linear state and consider its corresponding system matrix
  $-\vect{I} + \Sigmab \Wbf$. Using the same decomposition as
  in~\eqref{eq:W-decomp-P}, the system eigenvalues are $-1$ with
  multiplicity $N - r$ and the eigenvalues of the $r \times r$-matrix
  $\vect{P}$. For the latter, consider the characteristic polynomial
  of $\vect{P}$,
    $\operatorname{Char}(\vect{P} - \lambda \vect{I} ) = (-1)^{r}
    \lambda^{r} + (-1)^{r-1} K_{r-1}\lambda^{r-1}
    + \dots + (-1)K_{1} \lambda + K_0$ ,
  where $K_k$ represents the sum of all the principal minors of order
  $r - k $. In particular, since $r > 2$, $ K_{r-2} = \sum
  \limits_{\substack{i \neq j \text{ with } \sigma_i, \sigma_j = l}} M_{i,j} .  $
  Since the network is pairwise unstable, we deduce $ K_{r-2} < 0$
  and, consequently, $\sign((-1)^{{r}}) \neq \sign((-1)^{{r}-2}
  K_{{r}- 2})$.  Given that the characteristic polynomial has a sign
  change in its coefficients, using the Routh-Hurwitz
  criteria~\citep{AH:95} we deduce that there exists a root $\lambda$
  of the characteristic polynomial with $\text{Re}(\lambda) > 0 $, as
  claimed. \qed
\end{pf}

The implication of Theorem~\ref{PW_I_FUN} is that the analysis of LoSE
for pairwise unstable inhibitory networks can be reduced to the study
of those switching regions where only up to one node is in linear
state.  This is what we do in our next result.

\begin{proposition}\longthmtitle{Characterization of LoSE in pairwise
  unstable networks}\label{prop:NCSI}
Let $\Wbf$ be a pairwise unstable fully
inhibitory network. Define
  \begin{align*}
    \Tbb_0 & = \setdefb{\vect{u}}{ \exists i \in \until{N} \text{ s.t. }
      u_i > 0} ,
    \\
    \Tbb_i & = \setdefb{\vect{u}}{\bigvee \limits_{i \neq j \in
        \until{n}}( u_j > \frac{d_{j,i}}{d_{i,i} + 1} u_i)},
  \end{align*}
  for $i \in \until{N}$, and let $\Tbb = \bigcap_{i \in \ontil{n}}
  \Tbb_i$.  For a given $\mbf$, and if $ \ubf \in C = [0,(d_{1,1} + 1)
  m_1 ) \times \dots \times [0, (d_{N,N}+ 1) m_N) \neq \emptyset$,
  then LoSE holds iff $\vect{u} \in \Tbb$.
\end{proposition}
\vspace*{-10pt}
\begin{pf}
  From Theorem~\ref{PW_I_FUN}, the equilibrium candidate of any
  switching region with more than one node in linear state is
  unstable.  In addition, one can show that no switching region with a
  node in positive saturation can contain its corresponding
  equilibrium candidate. This is because the dynamics for such node,
  say $k$, would take the form
  \begin{equation*}
    \tau \dot{x}_k = - x_k + [- \sum_{i \neq k} d_{k,i}x_i  -d_{k,k}m_k
    + u_k ]_0 ^{m_k} . 
  \end{equation*}
  Since $\vect{u} \in C$, we deduce $u_k < (d_{k,k} + 1)m_k$, and so $
  - d_{k,k}m_k + u_k < m_k$. Consequently, the node always goes out of
  positive saturation. Similarly, for the switching region where all
  nodes are in negative saturation, the fact that the equilibrium
  candidate falls outside it is a consequence of $\ubf \in \Tbb_0$.
  Finally, for the switching region where node $i \in \until{N}$ is in
  linear state and all others are in negative saturation, its
  corresponding equilibrium candidate falls outside it iff
  $\ubf \in \Tbb_i$. \qed
\end{pf}

Given Proposition~\ref{prop:NCSI}, we next focus on understanding the
conditions on the network connectivity matrix ensuring that
$\Tbb$ is nonempty.
We first note that such conditions must involve at least three
nodes. This is because if only two nodes, say $i$ and $j$, are
considered then, by the pairwise instability assumption,
$\frac{d_{j,i}}{d_{i,i} + 1} < \frac{d_{j,j} + 1}{d_{i,j}}$, and
therefore if $u_j > \frac{d_{j,i}}{d_{i,i} + 1} u_i$ then $u_i <
\frac{d_{i,j}}{d_{j,j} + 1} u_j$, and vice versa.  To find then
conditions involving three or more nodes, we
re-interpret the inequalities that define $\Tbb _{-0} := \bigcap_{i
  \in \until{N}} \Tbb_i$ using graph-theoretic concepts.  Consider the
weighted complete graph with vertex set $\until{N}$, edge set
$\until{N} \times \until{N} \setminus \setdef{(i,i)}{i \in \until{N}}$
(i.e., self-loops are excluded), and weight matrix
\begin{equation*}
  \Fbf = \begin{bmatrix}
    0 & \frac{d_{1,1} +1 }{d_{2,1}} & \frac{d_{1,1} +1 }{d_{3,1}}  &
    \dots & \frac{d_{1,1} +1 }{d_{N,1}}  
    \\ 
    \frac{d_{2,2} +1 }{d_{1,2}}\
    & 0 & \frac{d_{2,2} +1 }{d_{3,2}}  & \dots & \frac{d_{2,2} +1
    }{d_{N,2}}
    \\
    \frac{d_{3,3} +1 }{d_{1,3}} &\frac{d_{3,3} +1 }{d_{2,3}}   & 0 &
    \dots & \vdots 
    \\
    \vdots & \vdots &  \vdots& \ddots & \vdots
    \\
    \frac{d_{N,N} +1 }{d_{1,N}}& \frac{d_{N,N} +1 }{d_{2,N}}  &
    \frac{d_{3,3} +1 }{d_{N,3}} & \dots & 0 
  \end{bmatrix} .
\end{equation*}
In this definition, edge $(i,j)$ corresponds to the inequality $u_j
\frac{d_{i,i} +1}{d_{j,i}} > u_i$. In this way, the row $i$ of $\Fbf$
corresponds to the set of inequalities that define the set~$\Tbb_i$.
To find conditions such that $\Tbb_{-0}$ is not empty, it is necessary
and sufficient that there exists a path that involves every node and
corresponds to a feasible sequence of inequalities. Note that $\Tbb_i$
is not empty when some inequality holds, meaning that node~$i$ has an
outgoing edge. Then, for $\Tbb_{-0}$ to be non empty, every node needs
to have an outgoing edge.  This is only possible if a cycle exists,
restricting all those $u_i$ involved in~it.  For those~$i$ not
involved in the cycle, there always exists a sufficiently small value
of $u_i$ that ensures $\Tbb_i$, and consequently $\Tbb_{-0}$, is not
empty.

Given these observations, we consider the collection of cycles of
length $3$ or more of the graph defined above.  This collection
represents all the ways the inequalities involved in the definition of
the set $\Tbb_{-0}$ can be satisfied while remaining compatible with
the pairwise instability condition.
For each cycle $G_c = (V_c,E_c)$, consider the connectivity matrix
$\Fbf_c$, of dimension $\card{V_c}$, that results from having the
edges inherit their weights from the full adjacency matrix~$\Fbf$.
The matrix $\Fbf_c$ has one non-zero element per row and column.
Consequently, for the cycle defined by~$G_c$, we have successfully
reduced the feasibility problem of the inequalities to the problem of
finding $\vect{v}$ such that $\Fbf_c \vect{v} > \vect{v}$ holds
componentwise.  If $\vect{v}$ exists, then the inequalities defined by
$G_c$ are feasible, and the set $\Tbb_{-0}$ is not empty. Moreover, if
the resulting $\vect{v}$ has some positive component, then the set
$\Tbb$ is not empty.

\begin{theorem}\longthmtitle{Sufficient condition for LoSE in pairwise
    unstable networks}\label{Thm_Cycle}
  Let $\Wbf$ be a pairwise unstable fully inhibitory network. If there
  is cycle whose adjacency matrix satisfies $\rho(\Fbf_c) > 1$, then
  there exists $\ubf$ for which LoSE holds.
\end{theorem}
\vspace*{-10pt}
\begin{pf}
  Let $G_c$ be a cycle whose adjacency matrix $\Fbf_c$ satisfies
  $\rho(\Fbf_c) > 1$. Since $G_c$ is strongly connected, $\Fbf_c$ is
  irreducible.
  Using the Perron-Frobenius theorem for irreducible
  matrices~\citep[Theorem 1.11]{FB-JC-SM:08cor}, we deduce that
  $\rho(\Fbf_c)$ is an eigenvalue of $\Fbf_c$ and has an eigenvector
  $\vect{v}$ with positive components.  Since $\rho (\Fbf_c)> 1$,
  $\Fbf_c \vect{v} = \rho(\Fbf_c) \vect{v} > \vect{v}$ element-wise.
  We can use this eigenvector to construct $\ubf$ belonging to $\Tbb$
  and $C$ as follows.  Let $\lambda \in (0,\min_{i \in V_c}
  \frac{(d_{i,i} + 1)m_i\norm{\vbf}}{v_i})$. Then, for every $i$ in
  the cycle, let $ u_i = \frac{v_i}{\norm{\vect{v}}} \lambda$.  Since
  $\Fbf_c \vect{v} > \vect{v}$, we have $\ubf \in \Tbb_i$ for every
  $i$ in the cycle. Moreover, by definition, $u_i \le m_i
  (d_{i,i}+1)$.  Since the components of $\vect{v}$ are all positive,
  so are the ones of $\ubf$.  For those nodes $j $ that do not belong
  to the cycle, we can find values that satisfy the inequalities by
  setting $u_j=0$. Since the entries $u_i$ are positive for all the
  nodes $i$ in the cycle, the vector $\ubf$ so
  constructed belongs to $ \Tbb$ and $C$, and LoSE follows from
  Proposition~\ref{prop:NCSI}. \qed
\end{pf}

\vspace*{-2ex}
We refer to the cycle in Theorem~\ref{Thm_Cycle} as \emph{valid}. The
next result guarantees the necessity of the existence of a valid cycle
when the input is restricted to have small values.

\begin{corollary}\longthmtitle{Necessary condition for LoSE in
    pairwise unstable networks with  small inputs}
  Let $\Wbf$ be a pairwise unstable fully inhibitory network. If
  $\ubf$ is restricted to~$C$,  a valid cycle exists iff there is
  $\ubf$ for which LoSE~holds.
\end{corollary}
\begin{pf}
  The implication from left to right follows from
  Theorem~\ref{Thm_Cycle}. To show the other implication, by
  Proposition~\ref{prop:NCSI}, we only need to prove that, when
  $\vect{u} \in C$, the existence of a valid cycle is necessary for
  $\Tbb$ to be not empty.  We reason by contradiction, i.e., assume
  there does not exist any valid cycle but $\Tbb \neq \emptyset$. Let
  $\vect{u} \in C\cap\Tbb$. As $\vect{u} \in \Tbb$, there exists a
  feasible sequence of inequalities. Let $G_c$ be the corresponding
  cycle, say of $t$ nodes $i_1, \dots , i_ t$,  encoding this sequence,
  \begin{align*}
    u_{i_1} < u_{i_2} \frac{d_{i_1,i_1} +1}{d_{i_2,i_1}},
    \; \dots  , \;
    u_{i_t} < u_{i_1} \frac{d_{i_t,i_t} +1}{d_{i_1,i_t}}
  \end{align*}
  holds, which implies $ \frac{d_{i_1,i_1}
    +1}{d_{i_2,i_1}}\frac{d_{i_2,i_2} +1}{d_{i_3,i_2}} \dots
  \frac{d_{i_t,i_t} +1}{d_{i_1,i_t}}> 1$.  Due to the structure of the
  adjacency matrix $\Fbf_c$ of the cycle, this means that $\det
  (\Fbf_c)>1$, and hence $\rho(\Fbf_c) > 1$, implying that the cycle
  is valid, which is a contradiction.
\qed
\end{pf}

The graph-theoretical approach to characterize LoSE in pairwise
unstable inhibitory networks can also be used to derive conditions on
how the system oscillations occur.

\begin{theorem}\longthmtitle{Node outside valid cycle does not oscillate}
  Consider a pairwise unstable fully inhibitory network and let $i$ be
  one if its nodes.  There exists $ \vect{u} \in C$ that provides lack
  of stable equilibria for which the node $i$ does not oscillate,
  i.e. is always found in the same saturated state, if and only if the
  node $i$ does not belong to the valid cycle associated to~$\ubf$.
\end{theorem}
\vspace*{-15pt}
\begin{pf}
  We prove the implication from left to right (the other one can be
  reasoned analogously).  As the node $i$ does not oscillate, then it
  must be in positive or negative saturation. As $\ubf \in C$, $u_i <
  (d_{i,i} + 1)m_i$, then it must be in negative saturation, because
  the node cannot remain in positive saturation state. If a node is in
  negative saturation, then it does not contribute to the oscillations
  of the other nodes, meaning that it is effectively as considering a
  new network with $N-1$ nodes. For this network to oscillate for
  $\ubf_{-i}$, it is necessary that there exists a valid cycle (which
  will not include node~$i$). \qed
\end{pf}

\section{Oscillations in Networks of Networks}\label{sec:eis}

Here, we build on the results of Section~\ref{sec:single-network} to
study the oscillatory behavior of a network of oscillators, each
itself represented by a linear-threshold network.  Motivated by the
experimental and computational evidence in brain networks, we are
interested in the phenomena of synchronization and phase-amplitude
coupling.  Consider $n$ oscillators, each modeled by an E-I pair,
connected over a network with adjacency matrix $\Abf \in
\realnonneg^{n \times n}$ via their excitatory
nodes~\citep{SFM-FP-SG-MC-STG-JMV-DSB:16}. %
Since $\Abf$ captures inter-oscillator connections, its diagonal
entries are zero.  The dynamics of the resulting network of networks
is
\begin{subequations}\label{eq:blts}
  \begin{align}
    \Tbf \dot \xbf = -\xbf + [\Wbf \xbf + \ubf]_{\zeros}^\mbf,
  \end{align}
  where $ \xbf = [\xbf_1, \cdots , \xbf_n]$, $\xbf_i = [x_{i, 1},
  x_{i, 2}]$, $\ubf$ and $\mbf$ have similar decompositions, $ \Tbf =
  \diag(\tau_1, \tau_1, \tau_2, \tau_2, \dots, \tau_n, \tau_n) $, and
  \begin{align}
    \Wbf &= \diag(\Wbf_1, \dots, \Wbf_n) + \Abf \otimes \Ebf, \ \Ebf
    = \begin{bmatrix} 1 & 0 \\ 0 & 0 \end{bmatrix}\!,
    \\[-5pt]
    \Wbf_i &= \begin{bmatrix} a_i & -b_i \\ c_i & -d_i \end{bmatrix},
    \quad \ A_{ii} = 0, \quad \ i \in \until{n},
  \end{align}
\end{subequations}
and $\otimes$ denotes the Kronecker product.  We assume each E-I pair
oscillates on its own. The first question we address is whether the
pairs maintain oscillatory behavior once interconnected.

\begin{theorem}\longthmtitle{Excitatory-to-excitatory-coupled networks}\label{thm:eis} 
  Consider the dynamics~\eqref{eq:blts} and assume that each $\Wbf_i$
  satisfies the conditions of Theorem~\ref{thm:ei}. Then, the overall
  network does not have any stable equilibria if and only if
  \begin{align}\label{eq:eis}
    &\sum\nolimits_{j = 1}^N A_{ij} m_{j, 1} < \bar u_{i, 1} - u_{i, 1},
    \\
    \notag &\bar u_{i, 1} \triangleq b_i \min\Big\{m_{i, 2},
    \frac{u_{i, 2} + c_i m_{i, 1}}{d_i + 1}\Big\} - (a_i - 1) m_{i,
      1},
  \end{align}
  holds for at least one $i \in \until{n}$. Moreover, the state of any
  E-I pair for which~\eqref{eq:eis} holds may not converge to a fixed
  value (except for trivial solutions
    at unstable equilibria, if any) irrespective of the validity of~\eqref{eq:eis} for other
  pairs.
\end{theorem}
\vspace*{-10pt}
\begin{pf}
  Consider an arbitrary $\sigmab \in \zls^{2n}$ and let $L \subseteq
  \until{n}, |L| = r$ be the set of pairs whose respective switching
  region from $\sigmab$ is unstable (i.e., $\sigmab_i = (\ell, j), j
  \in \zls, i \in L$). Let $\Pib = \bar \Pib \otimes \Ibf_2$ be the
  permutation matrix that permutes the pairs such that these $r$ pairs
  are placed first. Then, $\Pib(-\Ibf + \Sigmab \Wbf) \Pib^T = [\Rbf, \star; \zeros, \Nbf]$ where
    $\Rbf = -\Ibf + \Sigmab_L (\diag(\{\Wbf_i\}_{i \in L}) + \Abf_L
    \otimes \Ebf)$, $\Nbf = -\Ibf + \Sigmab_{L^c} \diag(\{\Wbf_i\}_{i \in L^c})$,
  and $\Sigmab_L$ is the $2r \times
  2r$ principal submatrix of $\Sigmab$ consisting of rows and columns
  corresponding to the pairs in $L$. $\Abf_L$ and $\Sigmab_{L^c}$
  are defined similarly.  Therefore, the eigenvalues of $-\Ibf +
  \Sigmab \Wbf$ consist of those of $\Rbf$ and $\Nbf$.

  $\Nbf$ has $n - r$ eigenvalues equal to $-1$ and $n - r$ eigenvalues
  that equal $-1 - d_i$ or $-1$, depending on whether
  $\sigma_{i, 2} = \ell$ or not for each $i \in L^c$. On the other
  hand, if $r > 0$, then
  \begin{align*}
    \tr(\Rbf) &= \tr(-\Ibf + \Sigmab_L \diag(\{\Wbf_i\}_{i \in L}))
    \\
    &\ge \tr(-\Ibf + \diag(\{\Wbf_i\}_{i \in L})) = \sum\nolimits_{i = 1}^r a_i -
    d_i - 2 > 0.
  \end{align*}
  Thus, any switching region $\Omega_\sigmab$ is stable \emph{if
    and only if} %
  $\sigma_{i, 1} \neq \ell$ for all $i \in \until{n}$. To prove the
  sufficiency of~\eqref{eq:eis}, consider any stable
  $\Omega_\sigmab$. Then, if~\eqref{eq:eis} holds for even one
  $i$,
  \begin{align*}
    u_{i, 1} + \sum\nolimits_{j = 1}^n A_{ij} (\xbf^*_\sigmab)_{j, 1} \le u_{i,
      1} + \sum\nolimits_{j = 1}^n A_{ij} m_{j, 1} \stackrel{\eqref{eq:eis}}{<}
    \bar u_{i, 1},
  \end{align*}
  ensuring $\xbf^*_\sigmab \notin \Omega_\sigmab$ (by
  Theorem~\ref{thm:ei}) and the sufficiency
  of~\eqref{eq:eis}. Regarding the last statement of the theorem, note
  that for $\xbf_i$ to converge to a fixed value, $\sum_j A_{ij}
  \xbf_{j, 1}(t)$ must either also converge to a fixed value or be
  greater than or equal to $\bar u_{i, 1} - u_{i, 1}$ for sufficiently
  large $t$, both contradicting~\eqref{eq:eis}.

  To prove the necessity of~\eqref{eq:eis}, assume that it does not
  hold for any $i$ or, in other words, at least one of
  \begin{subequations}
    \begin{align}
      \label{eq:eisn1} u_{i, 1} \!+\! \sum_{j = 1}^N A_{ij} m_{j, 1}
      \!&>\! b_i m_{i, 2} - (a_i - 1) m_{i, 1},
      \\
      \label{eq:eisn2} u_{i, 1} \!+\! \sum_{j = 1}^N A_{ij} m_{j, 1}
      \!&>\!  \frac{b_i(u_{i, 2} \!+\! c_i m_{i, 1})}{d_i \!+\! 1}
      \!-\! (a_i \!-\! 1) m_{i, 1},
    \end{align}
  \end{subequations}
  holds for all $i \in \until{n}$. Now, define $\sigmab \in \zls^n$ by
  \begin{align*}
    \sigmab_i =
    \begin{cases}
      (\s, \s) & \text{if} \ u_{i, 2} \ge (d_i + 1) m_{i, 2} -
      c_i m_{i, 1},
      \\
      (\s, \ell) & \text{if} \ u_{i, 2} < (d_i + 1) m_{i, 2} -
      c_i m_{i, 1}.
    \end{cases}
  \end{align*}
  Note that~\eqref{eq:eisn2} implies~\eqref{eq:eisn1} if
  $u_{i, 2} \ge (d_i + 1) m_{i, 2} - c_i m_{i, 1}$
  and~\eqref{eq:eisn1} implies~\eqref{eq:eisn2} otherwise. Given that
  all the excitatory nodes are at saturation in $\sigmab$, it is not
  difficult to show that $\Omega_\sigmab$ (which is stable, by the
  reasoning above) contains its equilibrium, showing the
  necessity of~\eqref{eq:eis}. \qed
\end{pf}

The assumptions of Theorem~\ref{thm:eis} are consistent with the
observation that long-range connections between different brain
regions are almost exclusively excitatory. Nevertheless, it is
possible that these excitatory connections target both excitatory and
inhibitory populations in the receiving region. Therefore, a more
realistic scenario is where the inter-network coupling consists of
both excitatory-to-excitatory and excitatory-to-inhibitory
connections. This generality, however, comes at the price that
condition~\eqref{eq:eis} becomes only sufficient.

\begin{theorem}\longthmtitle{Excitatory-to-all-coupled
    networks}\label{thm:Nblteei}
  Consider the dynamics~\eqref{eq:blt} with
  \begin{align*}
    \Wbf = \diag(\Wbf_1, \dots, \Wbf_N) + \Abf^e \otimes \begin{bmatrix}
      1 & 0 \\ 0 & 0 \end{bmatrix} + \Abf^i \otimes \begin{bmatrix} 0 &
      0 \\ 1 & 0 \end{bmatrix},
  \end{align*} 
  where $\Abf^e, \Abf^i \in \realnonneg^{N \times N}$, $ \diag(\Abf^e)
  = \diag(\Abf^i) = \zeros$,
  \begin{align*}
    &\Wbf_i = \begin{bmatrix} a_i & -b_i \\ c_i & -d_i \end{bmatrix},
    \quad a_i, b_i, c_i, d_i > 0, \ \forall i \in \until{N},
  \end{align*} 
  and each $\Wbf_i$ satisfies the conditions of
  Theorem~\ref{thm:ei}. Then, this system does not have any stable
  equilibria if
  \begin{subequations}\label{eq:Nblt-condsEI}
    \begin{align}
      \label{eq:Nblt-condsEIa} &\sum_{j = 1}^N A_{ij}^e m_{j, 1} < b_i
      m_{i, 2} - (a_i - 1) m_{i, 1} - u_{i, 1}
      \\
      \label{eq:Nblt-condsEIb} &\sum_{j = 1}^N [(d_i + 1) A_{ij}^e -
      b_i A_{ij}^i]^+ m_{j, 1}
      \\
      \notag &\qquad < \big(b_i c_i \!-\! (a_i \!-\! 1)(d_i \!+\!
      1)\big) m_{i, 1} \!-\! (d_i \!+\! 1) u_{i, 1} \!+\! b_i u_{i, 2}
      \\
      \label{eq:Nblt-condsEIc} &\sum_{j = 1}^N [b_i A_{ij}^i \!-\!
      (d_i \!+\! 1) A_{ij}^e]^+ m_{j, 1} < (d_i \!+\! 1) u_{i, 1}
      \!-\! b_i u_{i, 2}
    \end{align} 
  \end{subequations}
  all hold for at least one $i \in \until{N}$.
\end{theorem}
\vspace*{-10pt}
\begin{pf}
Consider $\sigmab \in
  \{0, \ell, s\}^{2N}$ and let $0 \le n \le N$ be the number of pairs
  whose respective switching region from $\sigmab$ is unstable (i.e.,
  $(\ell, 0)$, $(\ell, \ell)$, $(\ell, s)$). Without loss of
  generality, let them be the first $n$ pairs. Then,
  \begin{align*}
    -\Ibf + \Sigmab \Wbf = \Pib \begin{bmatrix} \Bbf_1 & \star & \star \\
      \zeros & \Bbf_2 & \star \\ 
      \zeros & \zeros & -\Ibf \end{bmatrix} \Pib^T,
  \end{align*}
  where $ \Bbf_1 = -\Ibf - \Sigmab_{n+1:N}^i \diag(d_{n+1}, \dots,
  d_N)$ and $ \Bbf_2 = -\Ibf + \Sigmab_{1:n} (\diag(\Wbf_1, \dots,
  \Wbf_n) + \Abf_{1:n} \otimes \diag(1, 0))$, and $\Pib$ is a
  permutation matrix to separate the excitatory and inhibitory nodes
  of the stable pairs. Therefore, similar to Theorem~\ref{thm:eis},
  $\sigmab \in \{0, \ell, s\}^{2N}$ is stable if and only if all its
  $N$ subindices are stable. Assume this is the case
  and~\eqref{eq:Nblt-condsEI} holds (at least) for $i \in
  \until{N}$. Then, from~\eqref{eq:Nblt-condsEIa}, $ u_{i, 1} +
  \sum_{j = 1}^N A_{ij}^e \xbf_{j, 1}^* < b_i m_{i, 2} - (a_i - 1)
  m_{i, 1}$, and
  from~\eqref{eq:Nblt-condsEIb}-\eqref{eq:Nblt-condsEIc},
  \begin{align*}
    0 &< (d_i + 1) \Big(u_{i, 1} + \sum_{j = 1}^N A_{ij}^e \xbf_{j,
      1}^*\Big) - b_i \Big(u_{i, 2} + \sum_{j = 1}^N A_{ij}^i \xbf_{j,
      1}^*\Big)
    \\
    &< \big(b_i c_i - (a_i - 1)(d_i + 1)\big) m_{i, 1},
  \end{align*}
  ensuring that $\xbf_\sigmab^* \notin \Omega_\sigmab$. \qed
\end{pf}

Unlike Theorem~\ref{thm:eis}, the condition of
Theorem~\ref{thm:Nblteei} is not necessary. The reason is that even
if~\eqref{eq:Nblt-condsEI} is violated for all $i$, they need not be
violated with the same excitatory saturation patterns (i.e., vectors
in $\{0, s\}^N$ showing whether the excitatory node of each pair is in
negative or positive saturation) while in Theorem~\ref{thm:eis},
if~\eqref{eq:eis} is violated for any node, it would be with the
excitatory saturation pattern of $(s, \dots, s)$ (possibly among
others). This ensures the existence of at least one stable $\sigmab
\in \{0, \ell, s\}^N$ (whose excitatory elements are all $s$) that
contains its equilibrium candidate. On the other hand,
when~\eqref{eq:Nblt-condsEI} is violated for each $i$, it may be with
one or more excitatory activation patterns none of which may be shared
among all the pairs. Therefore, the necessary and sufficient condition
for lack of stable equilibria in this case is that the intersection of
the sets of excitatory activation patterns of all pairs is empty, with
the convention that this set is empty for any pair for
which~\eqref{eq:Nblt-condsEI} holds.

\section{Conclusions and Future Work}

We have studied nonlinear networked dynamical systems with bounded
linear-threshold activation functions and different classes of
architectures interconnecting excitatory and inhibitory nodes. Given
the arbitrary dimensionality of these networks, and motivated by the
Poincare-Bendixson theorem, we have relied on the lack of stable
equilibria (LoSE) as a system-based proxy for the commonly used
signal-based definitions of oscillatory dynamics.  Our main
contributions are various necessary and/or sufficient conditions on
the structure of linear-threshold networks for LoSE. In particular, we
considered three classes of network architectures motivated by
different aspects of mammalian cortical architecture: networks with
multiple excitatory and one inhibitory nodes, purely inhibitory
networks, and arbitrary networks of two-dimensional
excitatory-inhibitory subnetworks. Among the important avenues for
future work, we highlight the extension of our results to include
conduction delays, the robustness analysis to process noise, and the
characterization of phase-phase and phase-amplitude coupling.

\setcounter{section}{0}
\renewcommand{\thesection}{Appendix \Alph{section}}
\section{Lack of Stable Equilibria as a Proxy for Oscillations}\label{app:lose}
\renewcommand{\thesection}{\Alph{section}}

Throughout the paper, we employ LoSE as a proxy for oscillations, as
defined in Definition~\ref{def:osc}. Here we provide numerical
evidence that, at least for systems with linear-threshold dynamics,
this proxy is tight. The evidence is structured along three
directions. First, we perform a Monte Carlo sampling of a 10-node
linear-threshold network and show the strong overlap between networks
that satisfy Definition~\ref{def:osc} and those without stable
equilibria. Second, for the same sampled set, we perform a similar
comparison \emph{locally} around the boundaries of the LoSE parameter
set, and show that the transition from oscillating to non-oscillating
and the transition from LoSE to presence of stable equilibria are
tightly related. Third, we exploit the analytical characterizations in
Section~\ref{sec:eis} to show not only the tightness of LoSE as a
binary measure of the existence of oscillations, but also the
relationship between the distance of a network to the appearance of
stable equilibria and the strength of its oscillations.

To numerically measure the existence and strength of oscillations, we
construct an \emph{oscillation index} directly based on
Definition~\ref{def:osc}. First, we define a \emph{regularity index}
to quantify Definition~\ref{def:osc}(i), i.e., the existence of
distinct and pronounced resonance peaks in the power spectrum of a
state trajectory. After mean-centering all state trajectories
$x_i(t)$, we let $X_i(f)$ be the Fourier transform of $x_i(t)$, and
$f_i = \argmax_f |X_i(f)|$. The regularity index is defined as
\begin{align*}
  \chireg &= \max_{i = 1, \dots, n} \chi_{{\rm reg}, i}, \\
  \chi_{{\rm reg}, i} &= \frac{|X_i(f_i)|}{\max\{|X_i((1 - \epsilon)
    f_i)|, |X_i((1 + \epsilon) f_i)|\}} \!\in\! [1, \infty),
\end{align*}
where $\epsilon \in (0, 1)$. For each $i$, a value of $\chi_{{\rm
    reg}, i} = 1$ indicates a flat power spectrum (lack of
oscillations) whereas $\chi_{{\rm reg}, i} \to \infty$ indicates a
Dirac delta at $f_i$ (periodic oscillations). Clearly, the regularity
of oscillations lies on a continuum, with more regularity (less
chaotic behavior) as $\chi_{{\rm reg}, i}$ grows. We then take the
maximum of $\chi_{{\rm reg}, i}$ to obtain a regularity index of the
collection of state trajectories $\xbf(t)$.

Second, we quantify Definition~\ref{def:osc}(ii) (lack of a constant
asymptotic limit) simply by the steady state peak to peak amplitude of
the oscillating trajectories, normalized by its maximum value
possible, and maximized over all trajectories,
\begin{align*}
  \chipp = \max_{i = 1, \dots, n} \frac{\limsup_{t \to \infty} x_i(t)
    - \liminf_{t \to \infty} x_i(t)}{m_i}
\end{align*} 
The larger $\chipp$, the stronger the oscillations in (at least one
channel of) $\xbf(t)$, regardless of how regular or chaotic they
are. Inclusion of this second metric is critical in distinguishing
between oscillations that are extremely regular but almost vanishing
in magnitude (and hence devoid of any practical significance), and
oscillations with significant amplitudes.

We combine the regularity and peak to peak indices to obtain the
oscillation index,
\begin{align}\label{eq:chiosc}
  \chiosc = \chireg \cdot \chipp
\end{align} 
Among the various potential ways of combining $\chireg$ and $\chipp$,
this choice acts a \emph{conjunction} of regularity and strength
measures, so that a signal is considered oscillatory if it has high
regularity \emph{and} strength, as required in
Definition~\ref{def:osc}.

\subsection{Global Inspection via Monte-Carlo Sampling of Structural
  Parameters}\label{subsec:global}

We start our numerical inspection of the relationship between LoSE and
existence of oscillations using a global Monte-Carlo sampling of the
parameter space of linear-threshold networks.  In general, the
distribution of indices $\chireg$, $\chipp$, and
$\chiosc$ depend on the number and excitatory/inhibitory mix of the
nodes. However, this dependence is not critical while, at the same
time, sweeping over $N_E$ and $N_I$ would be computationally
prohibitive for our Monte-Carlo sampling. Therefore, we here generate
$20000$ random networks using the fixed medium-range values of $N_E =
N_I = 5$ and address the role of network size in
Section~\ref{subsec:eis}. We use parameter values drawn randomly and
independently from the following distributions
\begin{align*}
  &|w_{ij}| \sim \U(0, B), \quad u_i \sim \U(-B, B), \quad m_i \sim
  \U(1, B)
  \\
  &x_i(0) \sim \U(0, m_i), \quad \forall i, j = 1, \dots, n,
\end{align*} 
where $n = N_E + N_I = 10$ and $B = 10$ is an (arbitrary, but
necessary) upper bound on the parameter values. We employ the value of
$\tau = 1$ throughout as the timescale only compresses or stretches
the trajectories over time. For each random network, we first check
whether it possesses any stable equilibria from~\eqref{eq:eqc}. For
networks that lack any stable equilibria, we simulate their
trajectories, starting from random initial conditions, over a
sufficiently long time horizon%
\footnote{We simulate all network trajectories over $t \in [0, 2000]$
  with a time step of $0.01$ using MATLAB's \texttt{ode45} and use the
  final $5\%$ of the trajectories for the computation of $\chireg$ and
  $\chipp$.}  and compute their value of $\chiosc$
in~\eqref{eq:chiosc}. For networks that did have (one or more) stable
equilibria, we repeat the same but starting from $10$ different
initial conditions to capture the possibility of the co-existence of
oscillatory and equilibrium attractors.

\begin{figure}[tbh]
  \begin{tikzpicture}
    \node (pie) {\includegraphics[width=0.25\linewidth]{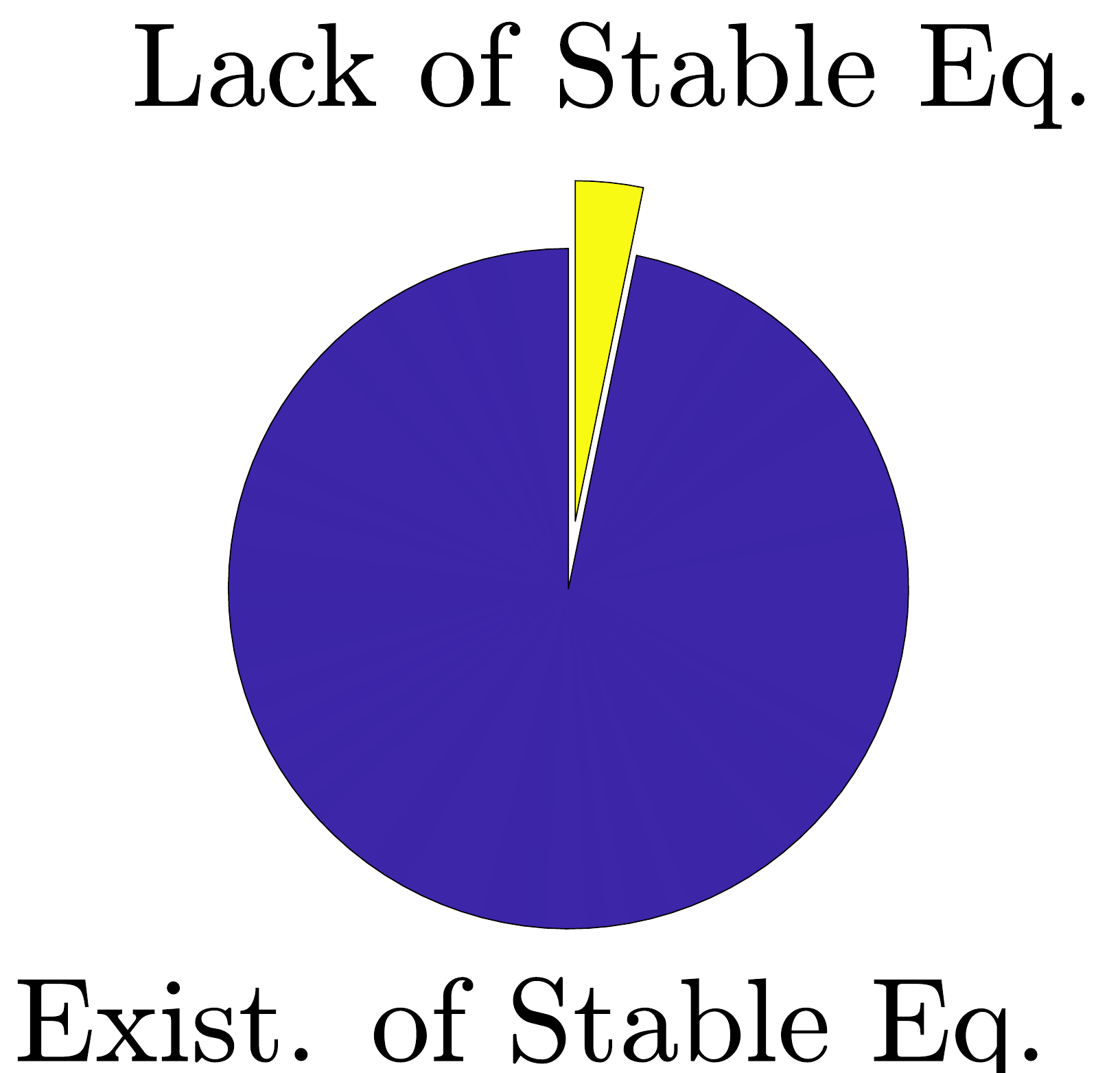}}; %
    \node[right of=pie, xshift=100pt, yshift=40pt] (lose) {\includegraphics[width=0.7\linewidth]{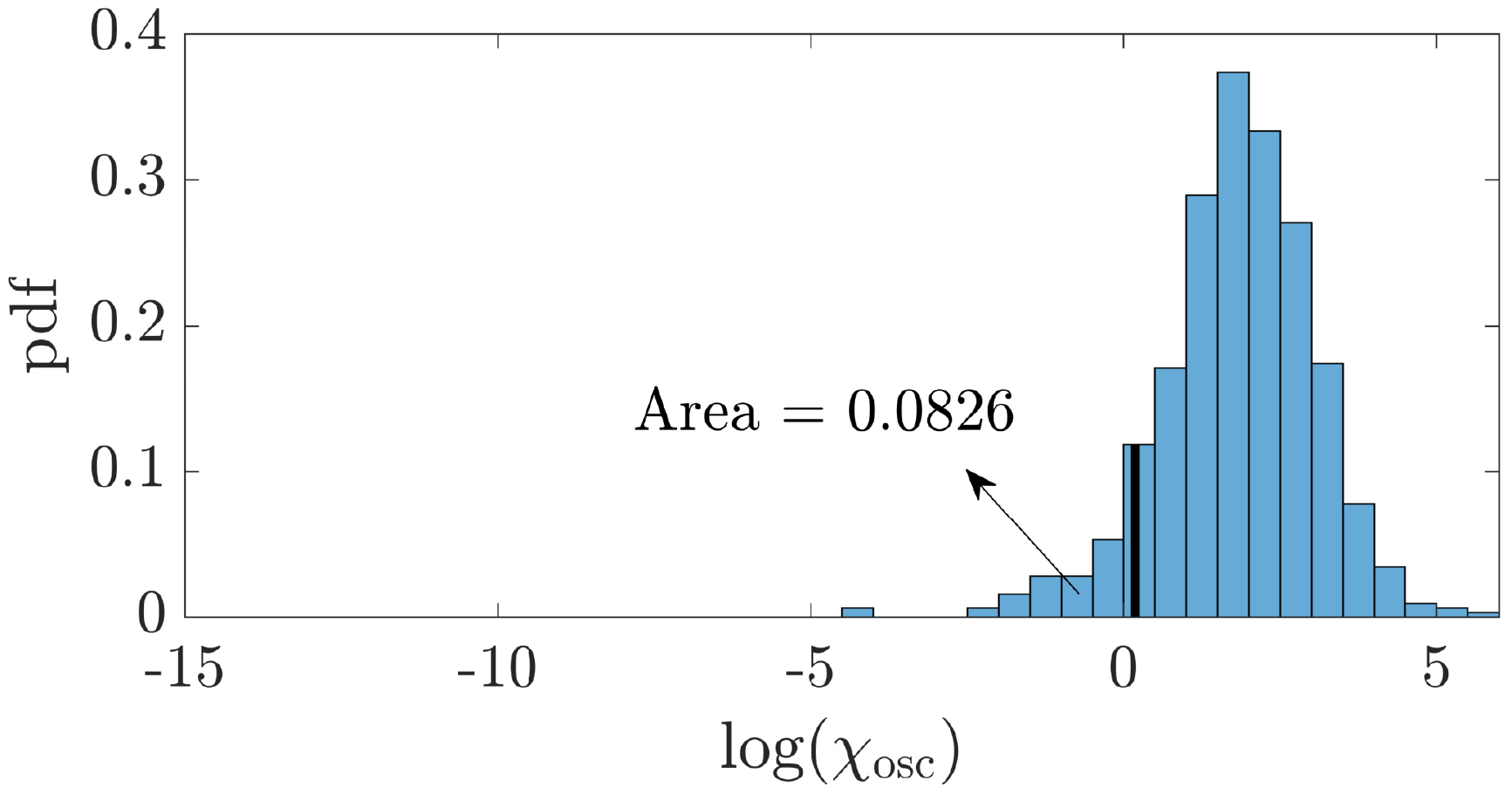}}; %
    \node[below of=lose, yshift=-70] (pose) {\includegraphics[width=0.7\linewidth]{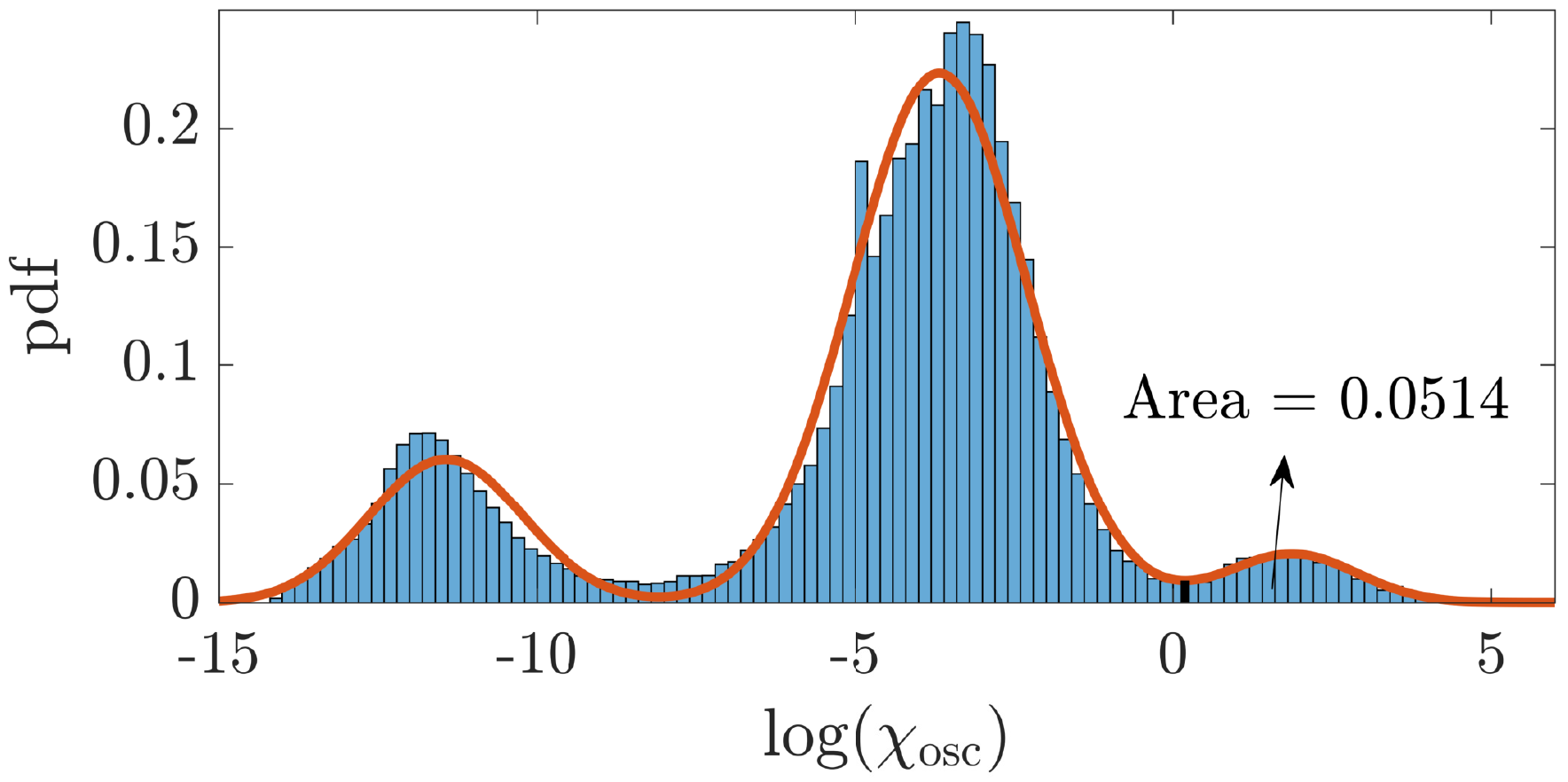}}; %
    \draw[-latex, bend left=35, shorten <=2pt, shorten >=5pt] (pie.north) to (lose.170); %
    \draw[-latex, bend right=35, shorten <=5pt, shorten >=5pt] (pie.south) to (pose.180); %
    \node[left of=lose, xshift=55pt, yshift=20pt, scale=0.7] (theta) {Threshold $\vartheta$}; %
    \draw[-latex, shorten >=1pt] (theta.-16) to +(0.5pt, -20pt); %
\end{tikzpicture} 
\vspace*{-10pt}
\caption{The statistics of LoSE and $\chiosc$ for randomly generated
  linear-threshold networks via Monte-Carlo sampling of their
  parameter space, as described in Section~\ref{subsec:global}.  Top
  right, only about $8\%$ of networks without stable equilibria lack
  strong oscillations (though the majority still possess weak
  oscillations). Bottom right, only about $5\%$ of networks with
  stable equilibria also have strongly oscillatory trajectories
  (corresponding to rare oscillatory attractors that co-exist with
  equilibrium attractors).  }
\label{fig:chires-rand}
\end{figure}
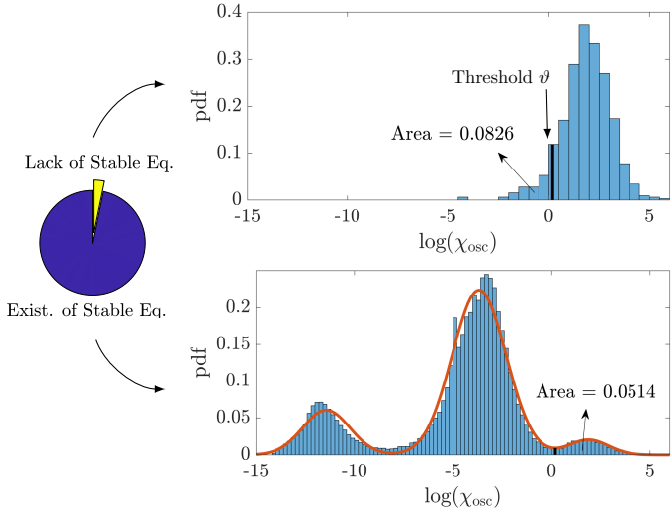

Figure~\ref{fig:chires-rand} shows the resulting statistics. First, we
observe that the lack of stable equilibria is less frequent than their
existence in random networks. Second, the values of $\chiosc$ lie on a
continuous spectrum, regardless of whether the networks possess or
lack stable equilibria. However, the distribution of $\chiosc$ is
significantly different between the two cases.

To quantify this difference, we need to place a threshold on the value
of $\chiosc$ and binarize the networks into ones that do show
oscillatory activity and ones that do not. In order to avoid using
arbitrary thresholds, we chose to obtain it from the empirical
distribution of $\chiosc$ we have just obtained. It can be seen from
the bottom-right panel of Figure~\ref{fig:chires-rand} that the
distribution of $\chiosc$ for networks with stable equilibria is
naturally tri-modal. The three chunks of the distribution correspond,
roughly, to strongly oscillating, barely oscillating, and effectively
non-oscillating trajectories, respectively. We thus fit a Gaussian
mixture model to this distribution and use the trough of the
distribution between the center and right modes as the threshold for
the existence of oscillations. Let this threshold be called
$\vartheta$. To ensure uniformity, $\vartheta$ is also used for
networks without stable equilibria.  Accordingly, we observe that only
about $8\%$ of networks \emph{without} stable equilibria lack strong
oscillations (though the majority of which still possess weak
oscillations) indicating the near-sufficiency of LoSE for existence of
oscillations. On the other hand, only about $5\%$ of networks
\emph{with} stable equilibria also have strongly oscillatory
trajectories (corresponding to rare oscillatory attractors that
co-exist with equilibrium attractors, each having their respective
regions of attraction) showing the near-necessity of LoSE for
exhibiting oscillations.

In conclusion, on a global landscape of the parameter space, LoSE
provides an unambiguous and system-based proxy with great analytical
utility for the existence of oscillations which closely matches the
signal-based definition of oscillations (cf. Definition~\ref{def:osc})
used in computational neuroscience.

\subsection{Local Inspection via Linear Sweeping of Structural
  Parameters}\label{subsec:local}

In this section, we assess the consistency of LoSE as a proxy for
oscillations on a local basis. Our basic idea is the following: given
a pair of networks, one which displays strong oscillations and another
that displays none, consider the convex combination of their
parameters $(\Wbf, \mbf, \ubf)$. As we traverse the resulting convex
set, the strong oscillations present on one extreme eventually
disappear into the non-oscillatory behavior of the other
extreme. Given our discussion above, the value of the convex parameter
where this transition occurs can be determined in two different ways:
either through LoSE or through the oscillatory metric $\chi_{\rm
  osc}$. The extent to which the two ways coincide offers a measure of
the \emph{local} consistency of LoSE as a proxy for oscillations.

\begin{figure}
\centering
\hspace{22.5pt}
\subfigure[]{\hspace{-15pt}\includegraphics[width=0.36\linewidth]{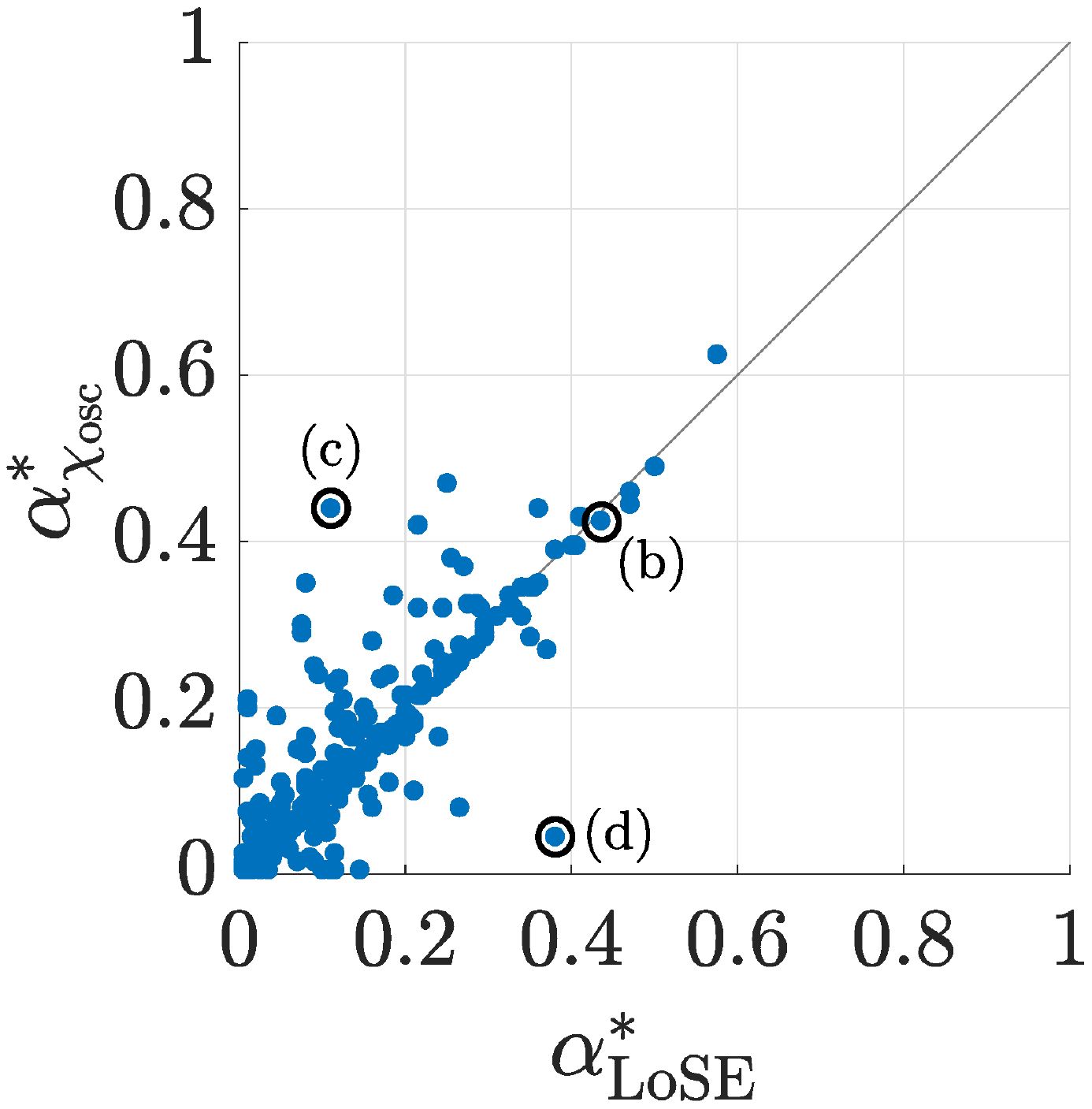}}
\hspace{22.5pt}
\subfigure[]{\includegraphics[width=0.45\linewidth]{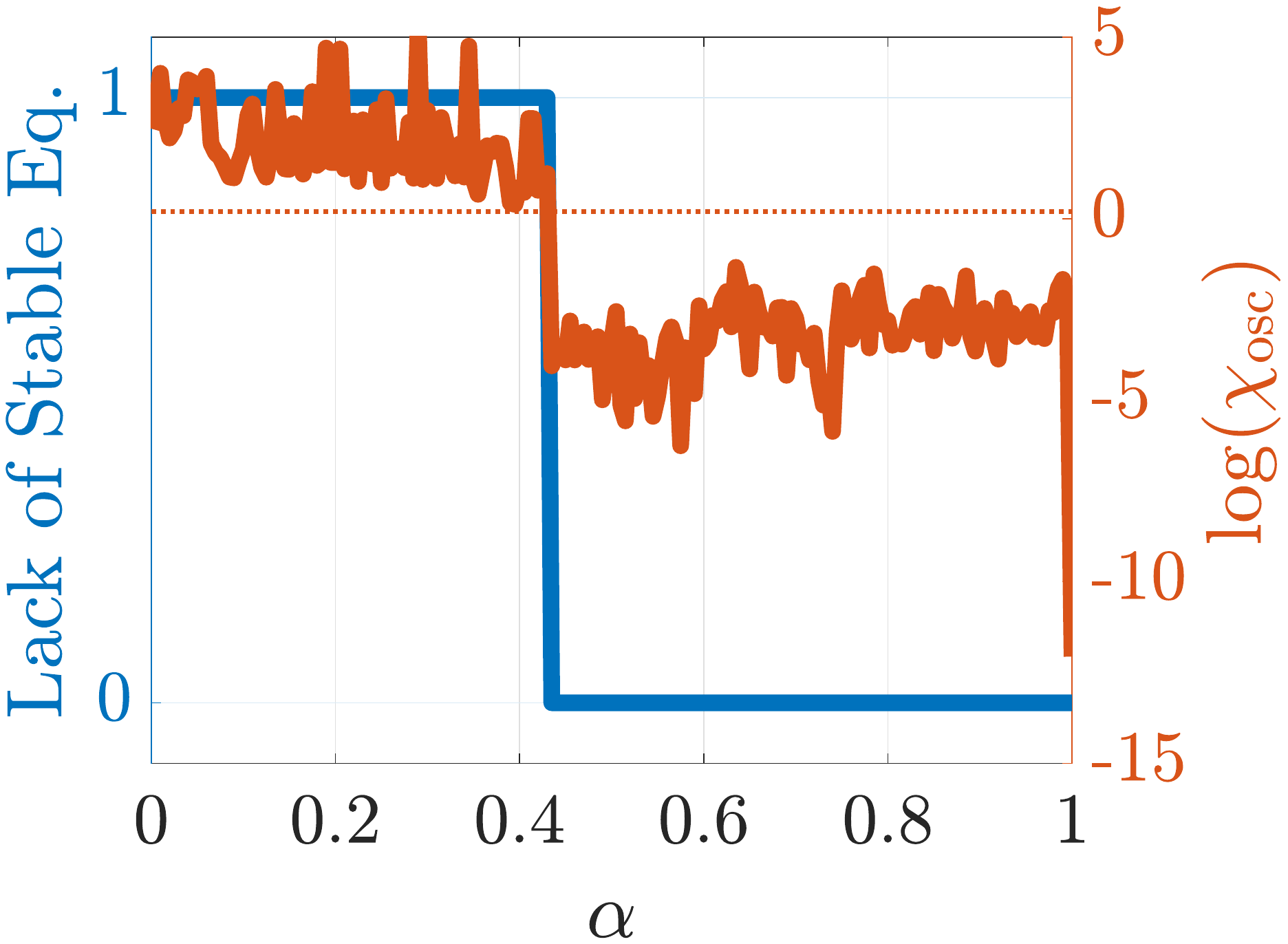}}
\\[-5pt]
\subfigure[]{\includegraphics[width=0.45\linewidth]{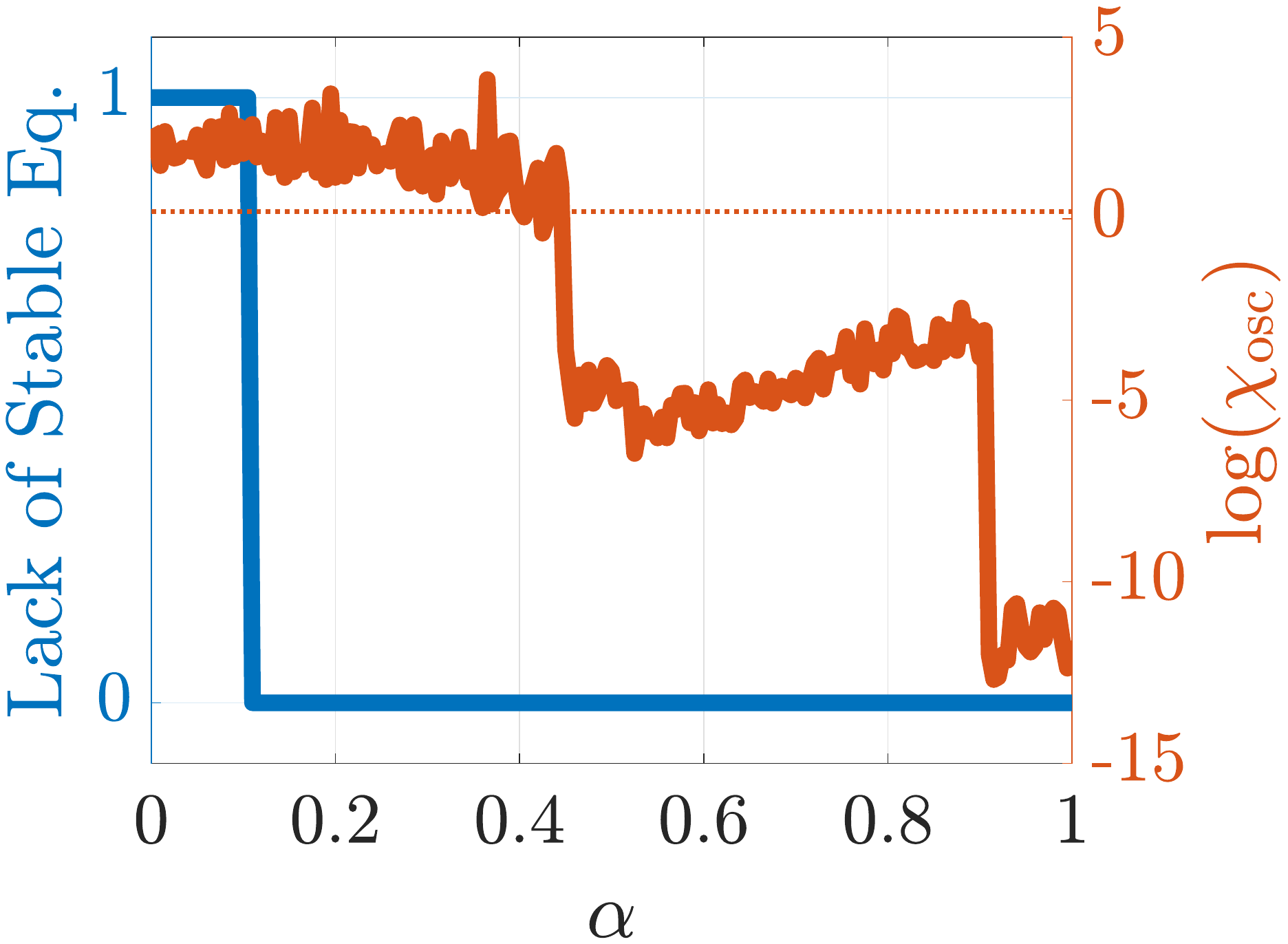}}
\subfigure[]{\includegraphics[width=0.45\linewidth]{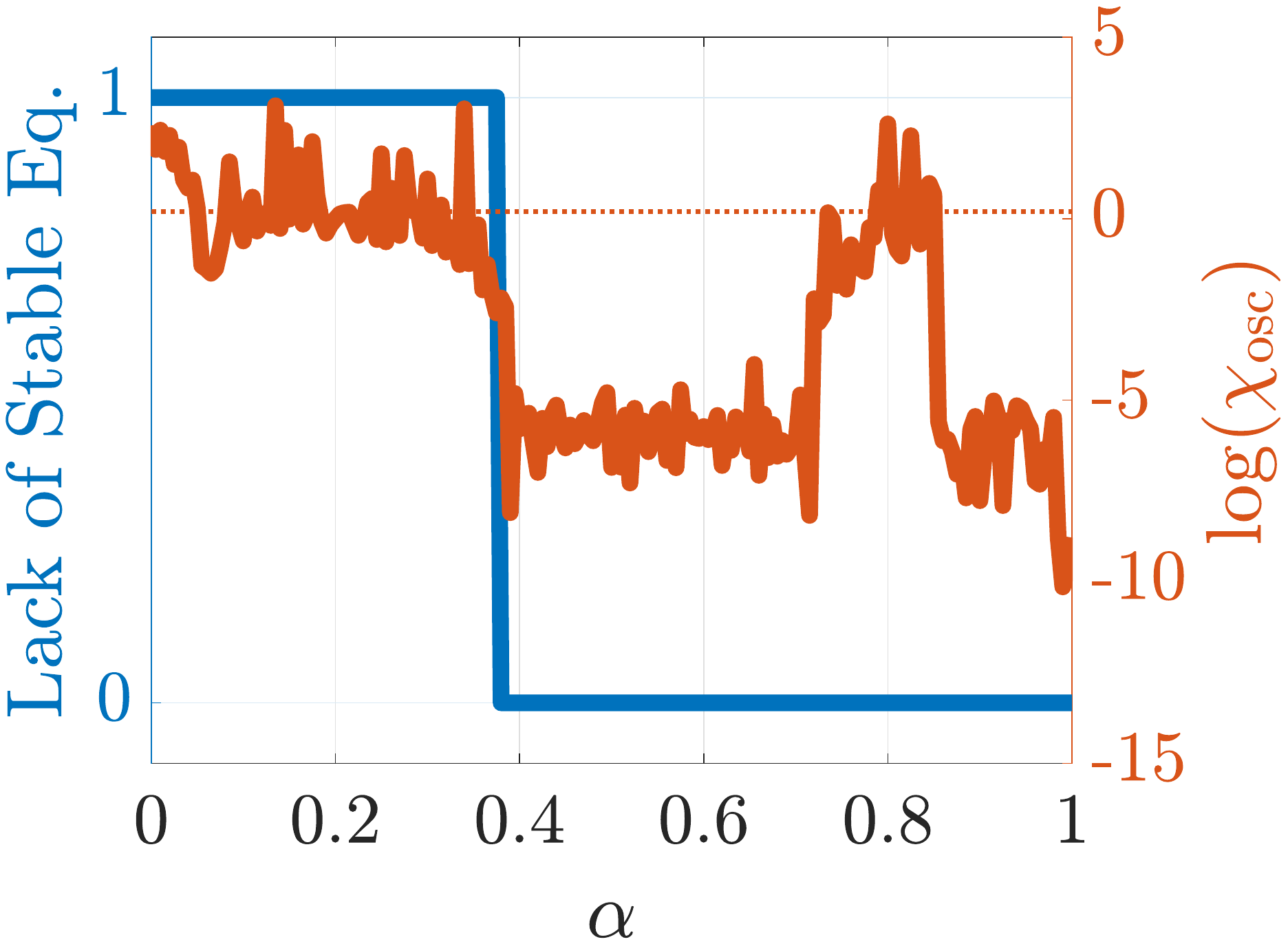}}
\vspace*{-10pt}
\caption{The consistency of LoSE (as a proxy for oscillations) and
  $\chi_{\rm osc}$ (as a ``ground truth" measure of oscillations) when
  locally sweeping between network parameters that give rise to
  oscillations and those that do not. \textbf{(a)} The value of
  $\alpha$ at which LoSE switches vs. the value of $\alpha$ at which
  $\log(\chi_{\rm osc})$ crosses the threshold $\vartheta$. 
  Note the gathering of the majority of
  the points around the $\alpha^*_{\chi_{\rm osc}} = \alpha^*_{\rm
    LoSE}$ line. \textbf{(b-d)} Sample plots of LoSE (left vertical
  axis) and $\log(\chi_{\rm osc})$ (right vertical axis) as a function
  of $\alpha$ for three sample cases denoted in panel (a). The red
  horizontal dotted line indicate the oscillation threshold $\vartheta$.
  Panel (b) illustrates a common
  mid-point scenario where $\alpha^*_{\chi_{\rm osc}} \simeq
  \alpha^*_{\rm LoSE}$ while (c) and (d) illustrate two extreme
  conditions.}
\label{fig:local}
\end{figure}

We carry out this vision by randomly selecting 500 pairs of networks
out of the 20000 generated in Section~\ref{subsec:global} as
follows. The first network of each pair is uniformly randomly selected
among the strongly oscillating networks of the top-right panel of
Figure~\ref{fig:chires-rand} (those to the right of the black vertical
line) that also lack stable equilibria, while the second network of
each pair is uniformly randomly selected from the almost
non-oscillating networks of the bottom-right panel of
Figure~\ref{fig:chires-rand} (those belonging to the left-most bump in
the distribution) that have some stable equilibria as well. Letting
$(\Wbf_1, \mbf_1, \ubf_1)$ and $(\Wbf_2, \mbf_2, \ubf_2)$ denote the
parameters of these networks, we then linearly sweep between the two
to obtain networks with parameters
  $\Wbf = (1 - \alpha) \Wbf_1 + \alpha \Wbf_2
  ,
  \mbf = (1 - \alpha) \mbf_1 + \alpha \mbf_2
  ,
  \ubf = (1 - \alpha) \ubf_1 + \alpha \ubf_2, \alpha \in [0, 1]$,
and compute LoSE and $\chi_{\rm osc}$ for each intermediate
network. Given the fact that the set of networks with LoSE is not
convex, we only retain the cases for which only one switching in LoSE
occurred between the end points as we sweep (due to the complexity of
estimating the switching point in $\chi_{\rm osc}$, as discussed
next). The value of $\alpha$ at which LoSE switches (i.e., a stable
equilibrium point appeared) is defined as~$\alpha^*_{\rm
  LoSE}$. Similarly, the value of $\alpha$ at which $\log(\chi_{\rm
  osc})$ crosses the threshold $\vartheta$ is defined as~$\alpha^*_{\chi_{\rm
    osc}}$. Due to the noisy nature of $\chi_{\rm osc}$ estimation
(see, e.g., Figure~\ref{fig:local}(b-d)), the numerical (or even
visual) detection of this threshold crossing is often not
straightforward. Here, we define $\alpha^*_{\chi_{\rm osc}}$ as the
first time (while increasing $\alpha$ from $0$ to $1$) that the
average of $3$ consecutive $\chi_{\rm osc}$ values is above
$\vartheta$ and the average of the following $3$ $\chi_{\rm osc}$ values
falls below $\vartheta$.

The resulting comparison of $\alpha^*_{\rm LoSE}$ and
$\alpha^*_{\chi_{\rm osc}}$ for the $500$ random pairs of networks
(except those having more than one switch in LoSE, as noted above) is
shown in Figure~\ref{fig:local}(a). Details of three sample scenarios
are also shown in Figure~\ref{fig:local}(b-d), with the corresponding
points marked in Figure~\ref{fig:local}(a). Even though not all the
points lie on the $\alpha^*_{\chi_{\rm osc}} = \alpha^*_{\rm LoSE}$
line, they are often very close to it, indicating a strong consistency
between the detection of oscillations using LoSE and $\chi_{\rm osc}$.

In addition to the closeness of the majority of the points to the
$\alpha^*_{\chi_{\rm osc}} = \alpha^*_{\rm LoSE}$ line, also notable
from Figure~\ref{fig:local}(a) is the fact that the majority of the
points lying away from this line lie above it, a situation exemplified
in Figure~\ref{fig:local}(c). This corresponds to scenarios where the
creation of the stable equilibrium point at $\alpha^*_{\rm LoSE}$ does
not immediately nullify the ongoing oscillatory attractor, but the two
coexist with distinct regions of attraction for some range of $\alpha$
values. The points lying below the $\alpha^*_{\chi_{\rm osc}} =
\alpha^*_{\rm LoSE}$ line, however, often indicate a complexity with
the detection of $\alpha^*_{\chi_{\rm osc}}$. An example of this can
be seen in Figure~\ref{fig:local}(d), where $\alpha^*_{\chi_{\rm
    osc}}$ is detected as the first threshold crossing, much sooner
(smaller) than $\alpha^*_{\rm LoSE}$, even though a meaningful drop in
$\chi_{\rm osc}$ is also clearly visible near $\alpha^*_{\rm
  LoSE}$. Note, also, that $\alpha^*_{\chi_{\rm osc}} < \alpha^*_{\rm
  LoSE}$ indicates a range of $\alpha$ values for which neither a
stable equilibrium point nor a strong oscillation exists. Since an
attractor must nevertheless exist, it can either be a highly chaotic
one (small $\chi_{\rm reg}$) or an oscillatory one with very small
amplitude (small $\chi_{\rm pp}$), neither of which we found to be
common in networks of size $n \simeq 10$.
   
\subsection{Global Inspection in Networks of E-I Pairs}\label{subsec:eis}

In Sections~\ref{subsec:global} and~\ref{subsec:local}, we have
inspected general excitatory-inhibitory networks with arbitrary
connection patterns between the nodes. Here, we inspect the networks
of E-I pairs studied in Section~\ref{sec:eis}. These networks not only
constitute an important special case from a computational neuroscience
standpoint, but they also lend themselves to theoretical
characterizations such as that in Theorem~\ref{thm:eis}. Here, we
inspect the quality of LoSE as a proxy for oscillations using the
theoretical condition in~\eqref{eq:eis}. To this end, we construct
random networks according to
\begin{align}\label{eq:rand}
  \notag &d_i \sim \U(0, d_{\max}), \quad a_i \sim \U(a_{\min},
  a_{\max}), \ a_{\min} > d_{\max} + 2,
  \\
  \notag &b_i = c_i \sim \U(b_{\min}, b_{\max}), \ b_{\min} >
  \sqrt{(a_{\max} - 1)(d_{\max} + 1)},
  \\
  \notag &m_{j, i} \sim \U(m_{j, \min}, m_{j, \max}), \ m_{2, \min} >
  \frac{a_{\max} - 1}{b_{\min}} m_{1, \max},
  \\
  &\tau_i \sim \U(\tau_{\min}, \tau_{\max}), \ \ {\rm i.i.d.} \
  \forall j = 1, 2, i \in \until{n},
\end{align}
all satisfying~\eqref{eq:eia}-\eqref{eq:eic}. The values of $u_{i, 1}$
and $u_{i, 2}$ are chosen at the center of their respective ranges
in~\eqref{eq:eid}-\eqref{eq:eie} so that the E-I pairs  oscillate at
their maximum amplitude before interconnection. For $\Abf$, we first
generate a random $\Gbf \in \realnonneg^{n \times n}$ with zero
diagonal and set $ \Abf = \eta \bar \Abf$, $\bar \Abf = \diag(\bar
\ubf_1 - \ubf_1) \Gbf [\diag(\Gbf \ones_n) \diag(\mbf_1)]^{-1}$.
$\Abf$ then satisfies~\eqref{eq:eis} for all $i \in \until{n}$ iff
$\eta \in [0, 1)$.

\begin{figure}
\centering
    \includegraphics[width=0.6\linewidth]{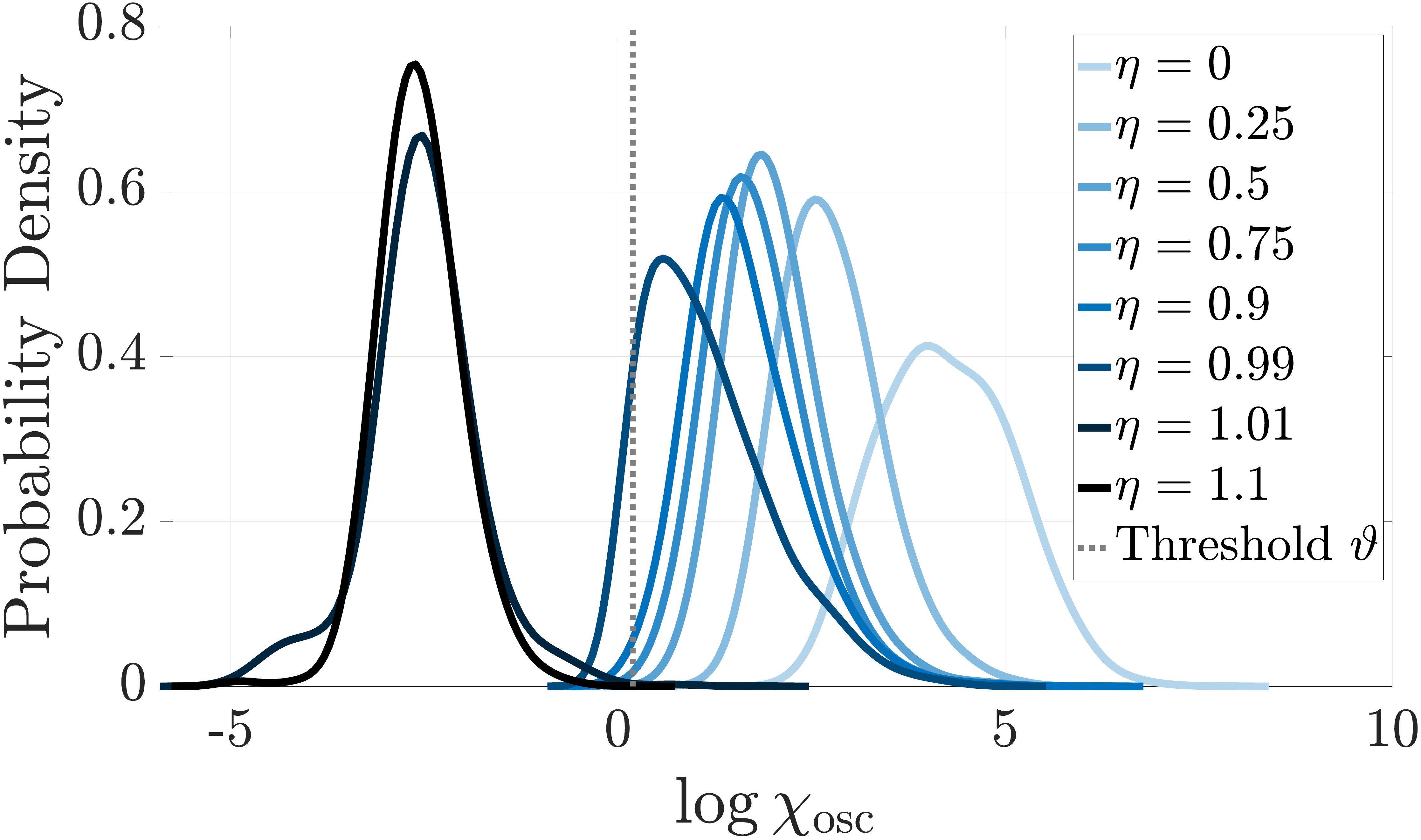}
\vspace*{-5pt}
  \caption{Strength and regularity of oscillations as a function of %
  inter-oscillator connection strength ($\eta$). The
    probability density function of $\log \chiosc$ is plotted
    for $n = 10$ and varying $\eta$. %
    Each distribution is based on $1000$
    random networks~\eqref{eq:rand} with $d_{\max} = 1$, $a_{\min} =
    3.5$, $a_{\max} = 5$, $b_{\min} = \sqrt{8} + 0.5$, $b_{\max} =
    \sqrt{8} + 2$, $m_{1, \min} = 1$, $m_{1, \max} = 2$, $m_{2, \min}
    = 8/b_{\min} + 0.5$, $m_{2, \max} = 8/b_{\min} + 2$, $\tau_{\min}
    \!=\! 1$, $\tau_{\max} \!=\! 10$.}
  \label{fig:res_ind}
  \vspace*{-1ex}
\end{figure}

Figure~\ref{fig:res_ind} shows the distribution of $\log \chiosc$ for
random networks of $n = 10$ oscillators, $\epsilon = 0.1$, and varying
interconnection strength~$\eta$. For disconnected oscillators $(\eta =
0)$, each oscillator has a perfectly regular oscillation (by
Theorem~\ref{thm:ei}) and thus very large $\chiosc$ (though finite,
due to the finiteness of $\chireg$, which is in turn due to finite
signal length and numerical error). These oscillations lose their
regularity and/or strength as we increase the connection strength
$\eta$ towards $1$, but still persist up to $\eta = 0.99$, showing the
almost sufficiency of~\eqref{eq:eis}. Moving beyond $\eta = 1$, almost
no oscillations persist even at $\eta = 1.01$ (and less so at $\eta =
1.1$) due to convergence to the stable equilibria ensured by
Theorem~\ref{thm:eis}. This shows that~\eqref{eq:eis} is also almost
necessary for existence of oscillations in the network
dynamics~\eqref{eq:blts}.

\section*{Acknowledgments}
The work was supported by NSF Award CMMI-1826065 (EN and JC) and ARO
Award W911NF-18-1-0213 (JC).  RP's stay at San Diego was funded by
the Centro de Formaci\'{o}n Interdisciplinaria Superior from
Universitat Polit\'{e}cnica de Catalu\~{n}a.


{
\small
\begin{thebibliography}{61}
\providecommand{\natexlab}[1]{#1}
\providecommand{\url}[1]{\texttt{#1}}
\expandafter\ifx\csname urlstyle\endcsname\relax
  \providecommand{\doi}[1]{doi: #1}\else
  \providecommand{\doi}{doi: \begingroup \urlstyle{rm}\Url}\fi

\bibitem[Angeli and Sontag(2003)]{DA-EDS:03}
David Angeli and Eduardo~D Sontag.
\newblock Monotone control systems.
\newblock \emph{IEEE Transactions on Automatic Control}, 48\penalty0
  (10):\penalty0 1684--1698, 2003.

\bibitem[Baird(1986)]{BB:86}
B.~Baird.
\newblock Nonlinear dynamics of pattern formation and pattern recognition in
  the rabbit olfactory bulb.
\newblock \emph{Physica D: Nonlinear Phenomena}, 22\penalty0 (1-3):\penalty0
  150--175, 1986.

\bibitem[Berger(1929)]{HB:29}
H.~Berger.
\newblock {\"U}ber das elektrenkephalogramm des menschen.
\newblock \emph{Archiv f{\"u}r Psychiatrie und Nervenkrankheiten}, 87\penalty0
  (1):\penalty0 527--570, Dec 1929.

\bibitem[Borisyuk and Kirillov(1992)]{RMB-ABK:92}
R.~M. Borisyuk and A.~B. Kirillov.
\newblock Bifurcation analysis of a neural network model.
\newblock \emph{Biological Cybernetics}, 66\penalty0 (4):\penalty0 319--325,
  1992.

\bibitem[Breakspear et~al.(2010)Breakspear, Heitmann, and
  Daffertshofer]{MB-SH-AD:10}
M.~Breakspear, S.~Heitmann, and A.~Daffertshofer.
\newblock Generative models of cortical oscillations: neurobiological
  implications of the {K}uramoto model.
\newblock \emph{Frontiers in Human Neuroscience}, 4:\penalty0 190, 2010.

\bibitem[Breakspear et~al.(2006)Breakspear, Roberts, Terry, Rodrigues, Mahant,
  and Robinson]{MB-JAR-JRT-SR-NM-PAR:06}
Michael Breakspear, JA~Roberts, John~R Terry, Serafim Rodrigues, N~Mahant, and
  PA~Robinson.
\newblock A unifying explanation of primary generalized seizures through
  nonlinear brain modeling and bifurcation analysis.
\newblock \emph{Cerebral Cortex}, 16\penalty0 (9):\penalty0 1296--1313, 2006.

\bibitem[Brouwer(1911)]{LEJB:11}
L.~E.~J. Brouwer.
\newblock {\"U}ber abbildung von mannigfaltigkeiten.
\newblock \emph{Mathematische Annalen}, 71\penalty0 (1):\penalty0 97--115,
  1911.

\bibitem[Bullo et~al.(2009)Bullo, Cort{\'e}s, and Martinez]{FB-JC-SM:08cor}
F.~Bullo, J.~Cort{\'e}s, and S.~Martinez.
\newblock \emph{Distributed Control of Robotic Networks}.
\newblock Applied Mathematics Series. Princeton University Press, 2009.
\newblock ISBN 978-0-691-14195-4.

\bibitem[Buzs{\'a}ki and Draguhn(2004)]{GB-AD:04}
G.~Buzs{\'a}ki and A.~Draguhn.
\newblock Neuronal oscillations in cortical networks.
\newblock \emph{Science}, 304\penalty0 (5679):\penalty0 1926--1929, 2004.

\bibitem[Buzsaki(2006)]{GB:06}
Gyorgy Buzsaki.
\newblock \emph{Rhythms of the Brain}.
\newblock Oxford University Press, 2006.

\bibitem[Campbell and Wang(1996)]{SC-DW:96}
S.~Campbell and D.~Wang.
\newblock Synchronization and desynchronization in a network of locally coupled
  {W}ilson-{C}owan oscillators.
\newblock \emph{IEEE Transactions on Neural Networks}, 7\penalty0 (3):\penalty0
  541--554, 1996.

\bibitem[Celi et~al.(2021)Celi, Allibhoy, Pasqualetti, and
  Cort\'es]{FC-AA-FP-JC:21-csl}
F.~Celi, A.~Allibhoy, F.~Pasqualetti, and J.~Cort\'es.
\newblock Linear-threshold dynamics for the study of epileptic events.
\newblock \emph{IEEE Control Systems Letters}, 5\penalty0 (4):\penalty0
  1405--1410, 2021.

\bibitem[Cole and Voytek(2017)]{SRC-BV:17}
S.~R. Cole and B.~Voytek.
\newblock Brain oscillations and the importance of waveform shape.
\newblock \emph{Trends in cognitive sciences}, 21\penalty0 (2):\penalty0
  137--149, 2017.

\bibitem[Cowan et~al.(2016)Cowan, Neuman, and van Drongelen]{JDC-JN-WvD:16}
J.~D. Cowan, J.~Neuman, and W.~van Drongelen.
\newblock {W}ilson-{C}owan equations for neocortical dynamics.
\newblock \emph{The Journal of Mathematical Neuroscience}, 6\penalty0
  (1):\penalty0 1, 2016.

\bibitem[Dayan and Abbott(2001)]{PD-LFA:01}
P.~Dayan and L.~F. Abbott.
\newblock \emph{Theoretical Neuroscience: Computational and Mathematical
  Modeling of Neural Systems}.
\newblock Computational Neuroscience. MIT Press, Cambridge, MA, 2001.

\bibitem[Destexhe and Sejnowski(2009)]{AD-TJS:09}
A.~Destexhe and T.~J. Sejnowski.
\newblock The {W}ilson-{C}owan model, 36 years later.
\newblock \emph{Biological Cybernetics}, 101\penalty0 (1):\penalty0 1--2, 2009.

\bibitem[Donoghue et~al.(2020)Donoghue, Haller, Peterson, Varma, Sebastian,
  Gao, Noto, Lara, Wallis, Knight, Shestyuk, and
  Voytek]{TD-MH-EJP-PV-PS-RG-TN-AHL-JDW-RTK-AS-BV:20}
Thomas Donoghue, Matar Haller, Erik~J Peterson, Paroma Varma, Priyadarshini
  Sebastian, Richard Gao, Torben Noto, Antonio~H Lara, Joni~D Wallis, Robert~T
  Knight, Avgusta Shestyuk, and Bradley Voytek.
\newblock Parameterizing neural power spectra into periodic and aperiodic
  components.
\newblock \emph{Nature Neuroscience}, 23\penalty0 (12):\penalty0 1655--1665,
  2020.

\bibitem[Ermentrout and Kopell(1990)]{GBE-NK:90}
G.~B. Ermentrout and N.~Kopell.
\newblock Oscillator death in systems of coupled neural oscillators.
\newblock \emph{SIAM Journal on Applied Mathematics}, 50\penalty0 (1):\penalty0
  125--146, 1990.

\bibitem[Ermentrout and Terman(2010)]{GBE-DHT:10}
G~Bard Ermentrout and David~H Terman.
\newblock \emph{Mathematical foundations of neuroscience}, volume~35.
\newblock Springer Science \& Business Media, 2010.

\bibitem[Fries(2015)]{PF:15}
P.~Fries.
\newblock Rhythms for cognition: Communication through coherence.
\newblock \emph{Neuron}, 88:\penalty0 220--235, 2015.

\bibitem[Grasman(1977)]{WG:77}
W.~Grasman.
\newblock Periodic solutions of autonomous differential equations in
  higher-dimensional spaces.
\newblock \emph{The Rocky Mountain Journal of Mathematics}, 7\penalty0
  (3):\penalty0 457--466, 1977.

\bibitem[Harris and Ermentrout(2015)]{JH-BE:15}
Jeremy Harris and Bard Ermentrout.
\newblock Bifurcations in the {W}ilson-{C}owan equations with nonsmooth firing
  rate.
\newblock \emph{SIAM Journal on Applied Dynamical Systems}, 14\penalty0
  (1):\penalty0 43--72, 2015.

\bibitem[Hirsch and Smith(2006)]{MWH-HS:06}
Morris~W Hirsch and Hal Smith.
\newblock Monotone dynamical systems.
\newblock In \emph{Handbook of differential equations: ordinary differential
  equations}, volume~2, pages 239--357. Elsevier, 2006.

\bibitem[Hopfield(1982)]{JJH:82}
J.~J. Hopfield.
\newblock Neural networks and physical systems with emergent collective
  computational abilities.
\newblock \emph{Proceedings of the National Academy of Sciences}, 79\penalty0
  (8):\penalty0 2554--2558, 1982.

\bibitem[Hoppensteadt and Izhikevich(2012)]{FCH-EMI:12}
Frank~C Hoppensteadt and Eugene~M Izhikevich.
\newblock \emph{Weakly connected neural networks}, volume 126.
\newblock Springer Science \& Business Media, 2012.

\bibitem[H{\"u}lsemann et~al.(2019)H{\"u}lsemann, Naumann, and
  Rasch]{MJH-EN-BR:19}
M.~J. H{\"u}lsemann, E.~Naumann, and B.~Rasch.
\newblock Quantification of phase-amplitude coupling in neuronal oscillations:
  Comparison of phase-locking value, mean vector length, modulation index, and
  generalized linear modeling cross-frequency coupling.
\newblock \emph{Frontiers in neuroscience}, 13:\penalty0 573, 2019.

\bibitem[Hurwitz(1895)]{AH:95}
A.~Hurwitz.
\newblock Ueber die bedingungen, unter welchen eine gleichung nur wurzeln mit
  negativen reellen theilen besitzt.
\newblock \emph{Mathematische Annalen}, 46\penalty0 (1):\penalty0 273--284,
  1895.

\bibitem[Inagaki et~al.(2019)Inagaki, Fontolan, Romani, and
  Svoboda]{HKI-LF-SR-KS:19}
H.~K. Inagaki, L.~Fontolan, S.~Romani, and K.~Svoboda.
\newblock Discrete attractor dynamics underlies persistent activity in the
  frontal cortex.
\newblock \emph{Nature}, 566\penalty0 (7743):\penalty0 212--217, 2019.

\bibitem[Izhikevich(2007)]{EMI:07}
E.~M. Izhikevich.
\newblock \emph{Dynamical Systems in Neuroscience}.
\newblock MIT press, 2007.

\bibitem[Jadi and Sejnowski(2014)]{MPJ-TJS:14}
M.~P. Jadi and T.~J. Sejnowski.
\newblock Regulating cortical oscillations in an inhibition-stabilized network.
\newblock \emph{Proceedings of the IEEE}, 102\penalty0 (5):\penalty0 830--842,
  2014.

\bibitem[Johansson(2003)]{MKJJ:03}
M.~K.~J. Johansson.
\newblock \emph{Piecewise Linear Control Systems: A Computational Approach}.
\newblock Lecture Notes in Control and Information Sciences. Springer Berlin
  Heidelberg, 2003.

\bibitem[Jones(2016)]{SRJ:16}
S.~R. Jones.
\newblock When brain rhythms aren't `rhythmic': implication for their
  mechanisms and meaning.
\newblock \emph{Current Opinion in Neurobiology}, 40:\penalty0 72--80, 2016.

\bibitem[Kalitzin et~al.(2019)Kalitzin, Petkov, Suffczynski, Grigorovsky,
  Bardakjian, da~Silva, and Carlen]{SK-GP-PS-VG-GLB-FL-PLC:19}
S.~Kalitzin, G.~Petkov, P.~Suffczynski, V.~Grigorovsky, B.~L. Bardakjian,
  F.~Lopes da~Silva, and P.~L. Carlen.
\newblock Epilepsy as a manifestation of a multistate network of oscillatory
  systems.
\newblock \emph{Neurobiology of Disease}, 130:\penalty0 104488, 2019.

\bibitem[Kissinger et~al.(2018)Kissinger, Pak, Tang, Masmanidis, and
  Chubykin]{STK-AP-YT-SCM-AAC:18}
S.~T. Kissinger, A.~Pak, Y.~Tang, S.~C. Masmanidis, and A.~A. Chubykin.
\newblock Oscillatory encoding of visual stimulus familiarity.
\newblock \emph{Journal of Neuroscience}, 38\penalty0 (27):\penalty0
  6223--6240, 2018.

\bibitem[Liberzon(2003)]{DL:03}
D.~Liberzon.
\newblock \emph{Switching in Systems and Control}.
\newblock Systems \& Control: Foundations \& Applications. Birkh{\"a}user,
  2003.

\bibitem[Mattia et~al.(2013)Mattia, Pani, Mirabella, Costa, Giudice, and
  Ferraina]{MM-PP-GM-SC-PD-SF:13}
M.~Mattia, P.~Pani, G.~Mirabella, S.~Costa, P.~Del Giudice, and S.~Ferraina.
\newblock Heterogeneous attractor cell assemblies for motor planning in
  premotor cortex.
\newblock \emph{Journal of Neuroscience}, 33\penalty0 (27):\penalty0
  11155--11168, 2013.

\bibitem[Menara et~al.(2020)Menara, Baggio, Bassett, and
  Pasqualetti]{TM-GB-DSB-FP:18}
T.~Menara, G.~Baggio, D.~S. Bassett, and F.~Pasqualetti.
\newblock Stability conditions for cluster synchronization in networks of
  heterogeneous {K}uramoto oscillators.
\newblock \emph{IEEE Transactions on Control of Network Systems}, 7\penalty0
  (1):\penalty0 302--314, 2020.

\bibitem[Monteiro et~al.(2002)Monteiro, Bussab, and Berlinck]{LHAM-MAB-JGCB:02}
L.~H.~A. Monteiro, M.~A. Bussab, and J.~G.~C. Berlinck.
\newblock Analytical results on a {W}ilson-{C}owan neuronal network modified
  model.
\newblock \emph{Journal of Theoretical Biology}, 219\penalty0 (1):\penalty0
  83--91, 2002.

\bibitem[Morrison et~al.(2016)Morrison, Degeratu, Itskov, and
  Curto]{KM-AD-VI-CC:16}
K.~Morrison, A.~Degeratu, V.~Itskov, and C.~Curto.
\newblock Diversity of emergent dynamics in competitive threshold-linear
  networks: a preliminary report.
\newblock \emph{arXiv preprint arXiv:1605.04463}, 2016.

\bibitem[Muldoon et~al.(2016)Muldoon, Pasqualetti, Gu, Cieslak, Grafton,
  Vettel, and Bassett]{SFM-FP-SG-MC-STG-JMV-DSB:16}
S.~F. Muldoon, F.~Pasqualetti, S.~Gu, M.~Cieslak, S.~T. Grafton, J.~M. Vettel,
  and D.~S. Bassett.
\newblock Stimulation-based control of dynamic brain networks.
\newblock \emph{PLOS Computational Biology}, 12\penalty0 (9):\penalty0
  e1005076, 2016.

\bibitem[Nozari and Cort\'es(2019)]{EN-JC:19-acc}
E.~Nozari and J.~Cort\'es.
\newblock Oscillations and coupling in interconnections of two-dimensional
  brain networks.
\newblock In \emph{{A}merican {C}ontrol {C}onference}, pages 193--198,
  Philadelphia, PA, July 2019.

\bibitem[Nozari and Cort\'es(2021)]{EN-JC:21-tacI}
E.~Nozari and J.~Cort\'es.
\newblock Hierarchical selective recruitment in linear-threshold brain
  networks. {P}art {I}: Intra-layer dynamics and selective inhibition.
\newblock \emph{IEEE Transactions on Automatic Control}, 66\penalty0
  (3):\penalty0 949--964, 2021.

\bibitem[Onslow et~al.(2014)Onslow, Jones, and Bogacz]{ACEO-MWJ-RB:14}
A.~C.~E. Onslow, M.~W. Jones, and R.~Bogacz.
\newblock A canonical circuit for generating phase-amplitude coupling.
\newblock \emph{PLOS One}, 9\penalty0 (8):\penalty0 e102591, 2014.

\bibitem[Papadopoulos et~al.(2020)Papadopoulos, Lynn, Battaglia, and
  Bassett]{LP-CWL-DB-DSB:20}
L.~Papadopoulos, C.~W. Lynn, D.~Battaglia, and D.~S. Bassett.
\newblock Relations between large-scale brain connectivity and effects of
  regional stimulation depend on collective dynamical state.
\newblock \emph{PLOS Computational Biology}, 16\penalty0 (9):\penalty0 1--43,
  09 2020.

\bibitem[Perko(2000)]{LP:00}
L.~Perko.
\newblock \emph{Differential Equations and Dynamical Systems}, volume~7 of
  \emph{Texts in Applied Mathematics}.
\newblock Springer, New York, 3rd edition, 2000.

\bibitem[Quentin et~al.(2019)Quentin, King, Sallard, Fishman, Thompson, Buch,
  and Cohen]{RQ-JK-ES-NF-RT-ERB-LGC:19}
R.~Quentin, J.~King, E.~Sallard, N.~Fishman, R.~Thompson, E.~R. Buch, and L.~G.
  Cohen.
\newblock Differential brain mechanisms of selection and maintenance of
  information during working memory.
\newblock \emph{Journal of Neuroscience}, 39\penalty0 (19):\penalty0
  3728--3740, 2019.

\bibitem[R{\u{a}}svan(2007)]{VIR:07}
Vl~R{\u{a}}svan.
\newblock A new dissipativity criterion—towards yakubovich oscillations.
\newblock \emph{International Journal of Robust and Nonlinear Control:
  IFAC-Affiliated Journal}, 17\penalty0 (5-6):\penalty0 483--495, 2007.

\bibitem[Sanchez(2010)]{LAS:10}
L.~A. Sanchez.
\newblock Existence of periodic orbits for high-dimensional autonomous systems.
\newblock \emph{Journal of Mathematical Analysis and Applications},
  363\penalty0 (2):\penalty0 409--418, 2010.

\bibitem[Sase et~al.(2017)Sase, Katori, Komuro, and Aihara]{TS-YK-MK-KA:17}
Takumi Sase, Yuichi Katori, Motomasa Komuro, and Kazuyuki Aihara.
\newblock Bifurcation analysis on phase-amplitude cross-frequency coupling in
  neural networks with dynamic synapses.
\newblock \emph{Frontiers in computational neuroscience}, 11:\penalty0 18,
  2017.

\bibitem[Schuster and Wagner(1990)]{HGS-PW:90}
H.~G. Schuster and P.~Wagner.
\newblock A model for neuronal oscillations in the visual cortex. 1. mean-field
  theory and derivation of the phase equations.
\newblock \emph{Biological Cybernetics}, 64\penalty0 (1):\penalty0 77--82,
  1990.

\bibitem[Segneri et~al.(2020)Segneri, Bi, Olmi, and Torcini]{MS-HB-SO-AT:20}
Marco Segneri, Hongjie Bi, Simona Olmi, and Alessandro Torcini.
\newblock Theta-nested gamma oscillations in next generation neural mass
  models.
\newblock \emph{Frontiers in computational neuroscience}, 14:\penalty0 47,
  2020.

\bibitem[Simic et~al.(2002)Simic, Johansson, Lygeros, and
  Sastry]{SS-KHJ-JL-SS:02}
S.~Simic, K.~H. Johansson, J.~Lygeros, and S.~Sastry.
\newblock Hybrid limit cycles and hybrid {P}oincar{\'e}-{B}endixson.
\newblock In \emph{{IFAC} {W}orld {C}ongress}, 2002.

\bibitem[Steriade(2006)]{MS:06-neuro}
M.~Steriade.
\newblock Grouping of brain rhythms in corticothalamic systems.
\newblock \emph{Neuroscience}, 137\penalty0 (4):\penalty0 1087--1106, 2006.

\bibitem[Tang et~al.(2007)Tang, Simsek, Ozdaglar, and
  Acemoglu]{AKT-AS-AO-DA:06}
A.~Kevin Tang, A.~Simsek, A.~Ozdaglar, and D.~Acemoglu.
\newblock On the stability of p-matrices.
\newblock \emph{Linear Algebra and its Applications}, 426\penalty0
  (1):\penalty0 22--32, 2007.

\bibitem[Tomberg and Yakubovich(1989)]{EAT-VAY:89}
EA~Tomberg and Vladimir~Andreevich Yakubovich.
\newblock Conditions for auto-oscillations in nonlinear systems.
\newblock \emph{Siberian Mathematical Journal}, 30\penalty0 (4):\penalty0
  641--653, 1989.

\bibitem[Tort-Colet et~al.(2019)Tort-Colet, Capone, Sanchez-Vives, and
  Mattia]{NT-CC-MVS-MM:19}
N.~Tort-Colet, C.~Capone, M.~V. Sanchez-Vives, and M.~Mattia.
\newblock Attractor competition enriches cortical dynamics during awakening
  from anesthesia.
\newblock \emph{bioRxiv}, 2019.
\newblock URL \url{https://www.biorxiv.org/content/early/2019/01/10/517102}.

\bibitem[van Ede et~al.(2018)van Ede, Quinn, Woolrich, and
  Nobre]{FvE-AJQ-MWW-ACN:18}
F.~van Ede, A.~J. Quinn, M.~W. Woolrich, and A.~C. Nobre.
\newblock Neural oscillations: sustained rhythms or transient burst-events?
\newblock \emph{Trends in Neurosciences}, 41\penalty0 (7):\penalty0 415--417,
  2018.

\bibitem[Wang(2010)]{XW:10}
X.~Wang.
\newblock Neurophysiological and computational principles of cortical rhythms
  in cognition.
\newblock \emph{Physiological Reviews}, 90\penalty0 (3):\penalty0 1195--1268,
  2010.

\bibitem[White et~al.(1995)White, Budde, and Kay]{JAW-TB-ARK:95}
John~A White, Thomas Budde, and Alan~R Kay.
\newblock A bifurcation analysis of neuronal subthreshold oscillations.
\newblock \emph{Biophysical Journal}, 69\penalty0 (4):\penalty0 1203--1217,
  1995.

\bibitem[Whittington et~al.(2000)Whittington, Traub, Kopell, Ermentrout, and
  Buhl]{MAW-RDT-NK-BE-EHB:00}
M.~A. Whittington, R.~D. Traub, N.~Kopell, B.~Ermentrout, and E.~H. Buhl.
\newblock Inhibition-based rhythms: experimental and mathematical observations
  on network dynamics.
\newblock \emph{International Journal of Psychophysiology}, 38\penalty0
  (3):\penalty0 315--336, 2000.

\bibitem[Wilson and Cowan(1972)]{HRW-JDC:72}
H.~R. Wilson and J.~D. Cowan.
\newblock Excitatory and inhibitory interactions in localized populations of
  model neurons.
\newblock \emph{Biophysical Journal}, 12\penalty0 (1):\penalty0 1--24, 1972.

\end{thebibliography}

}
\end{document}